\documentclass[12pt]{article}

\usepackage{color}
\definecolor{darkblue}{rgb}{0.1,0.1,.7}
\usepackage[colorlinks, linkcolor=darkblue, citecolor=darkblue, urlcolor=darkblue, linktocpage]{hyperref}
\usepackage[]{amsmath}
\usepackage[]{graphicx}
\usepackage[]{latexsym}
\usepackage{slashed,graphicx,color,amsmath,amssymb}
\usepackage[mathscr]{eucal}
\usepackage{mathrsfs}
\usepackage{geometry}
\usepackage[margin=10pt,font=small,labelfont=bf]{caption}
\usepackage{amscd}
\usepackage{bm}
\usepackage{xcolor}
\usepackage{upgreek}
\usepackage[square, comma, sort&compress,numbers]{natbib}
\usepackage[all,cmtip]{xy}
\numberwithin{equation}{section}
\newcommand{\tr}{\mathrm{Tr}\,}

\newcommand{\rs}{\mathrm{s}}

\newcommand{\rA}{\mathrm{A}}
\newcommand{\rB}{\mathrm{B}}
\newcommand{\rC}{\mathrm{C}}

\newcommand{\rF}{\mathrm{F}}
\newcommand{\rG}{\mathrm{G}}

\newcommand{\rK}{\mathrm{K}}

\newcommand{\rR}{\mathrm{R}}
\newcommand{\rS}{\mathrm{S}}

\newcommand{\LT}{\mathrm{LT}}

\begin{document}
\vspace*{-.6in} \thispagestyle{empty}
\vspace{.2in} {\Large
\begin{center}
{\bf Algebraic geometry and Bethe ansatz (I)\\ The quotient ring for BAE}
\end{center}}
\vspace{.2in}
\begin{center}
Yunfeng Jiang,  Yang Zhang
\\
\vspace{.3in}
\small{\textit{Institut f{\"u}r Theoretische Physik,
ETH Z{\"u}rich}},\\
\small{\textit{
Wolfgang Pauli Strasse 27,
CH-8093 Z{\"u}rich, Switzerland}
}

\end{center}

\vspace{.3in}

\begin{abstract}
\normalsize{In this paper and upcoming ones, we initiate a systematic study of Bethe ansatz equations for integrable models by modern computational algebraic geometry. We show that algebraic geometry provides a natural mathematical language and powerful tools for understanding the structure of solution space of Bethe ansatz equations. In particular, we find novel efficient methods to count the number of solutions of Bethe ansatz equations based on Gr\"obner basis and quotient ring. We also develop analytical approach based on companion matrix to perform the sum of on-shell quantities over all physical solutions without solving Bethe ansatz equations explicitly. To demonstrate the power of our method, we revisit the completeness problem of Bethe ansatz of Heisenberg spin chain, and calculate the sum rules of OPE coefficients in planar $\mathcal{N}=4$ super-Yang-Mills theory.

}
\end{abstract}

\vskip 1cm \hspace{0.7cm}

\newpage

\setcounter{page}{1}
\begingroup
\hypersetup{linkcolor=black}
\tableofcontents
\endgroup

\section{Introduction}
\label{sec:intro}
Bethe ansatz is a powerful tool to find exact solutions of integrable models. Ever since the seminal work of Hans Bethe \cite{Bethe:Anstaz}, the original method has been developed largely and the term `Bethe anatz' now refers to a whole family of methods with different adjectives such as \emph{coordinate} Bethe ansatz, \emph{algebraic} Bethe ansatz \cite{Korepin_book,Faddeev:ABA}, \emph{analytic} Bethe ansatz \cite{Reshetikhin1983} and \emph{off-diagonal} Bethe ansatz \cite{wang2015off:book}. A crucial step in the Bethe ansatz methods is to write down a set of algebraic equations called the Bethe ansatz equations (BAE). These equations can be derived from different point of views such as periodicity of the wavefunction, cancelation of `unwanted terms' and analyticity of the transfer matrix.\par

The BAE is a set of quantization conditions for the rapidities (or momenta) of excitations\footnote{It might also involve some auxiliary variables as in the case of integrable models with higher rank symmetry algebras.} of the model, the solutions of which are called Bethe roots. Physical quantities such as momentum and energy of the system are functions of the rapidities. Once the BAE is known, one can solve it to find the Bethe roots and plug into the physical quantities. Therefore, in many cases solving an integrable model basically means writing down a set of BAE for the model.\par

However, in many applications, simply writing down the BAE is not the end of the story. In fact, solving BAE is by no means a trivial task ! Due to the complexity of BAE, it can only be studied analytically in certain limits such as the thermodynamic limit \cite{Takahashi:stacks} and the Sutherland limit \cite{Sutherland} (or semi-classical limit \cite{Beisert:2003xu,Kazakov:2004qf}). In both cases the size of the system and the number of excitations are large or infinite. For finite system size and number of excitations, typically the BAE can only be solved numerically. While numerical methods are adequate for many applications in physics, they have their limits and shortcomings. Firstly, numerical solutions cannot give exact answers and one needs to find the solutions with high precisions to obtain reliable results. Also, numerical methods might suffer from additional subtleties such as numerical instabilities. Finally and most importantly, the algebraic structure and beauty of BAE can hardly be seen by solving the equations numerically.\par

From the mathematical point of view, BAE is a set of algebraic equations whose solutions are a collection of points in certain affine space and form a zero dimensional affine variety. It is therefore expected that algebraic geometry may play a useful role in studying the BAE. The first work in this direction was done by Langlands and Saint-Aubin who studied the BAE of six vertex model (or XXZ spin chain) using algebraic geometry \cite{langlands1995algebro}. Here we take a slightly different point of view and study BAE from the perspective of modern \emph{computational} algebraic geometry. In particular, we propose that Gr\"obner basis and quotient ring are the proper language to describe BAE. The aim of our current work is to initialize a more systematic study of the structure of BAE using the powerful tool of algebraic geometry and at the same time developing efficient methods to derive exact results which previously relies on solving BAE numerically.\par

To demonstrate our points, we study two types of problems with algebro-geometric methods. The first type of problem is a revisit of the completeness problem of Bethe ansatz. This is a longstanding problem for Bethe ansatz which will be discussed in more detail in section\,\ref{sec:completeness} and appendix\,\ref{app:BAEcompleteness}. Despite the general belief that the Bethe ansatz is complete and many non-trivial progress, this problem does not have a complete and satisfactory solution. In terms of BAE, the completeness problem amounts to counting the number of \emph{physical} solutions of BAE. Analytical formula for the number of solutions of BAE with various additional constraints in terms of quantum numbers\footnote{Such as the length of the spin chain and number of excitations.} are still unknown\footnote{By this we mean the number of all solutions with pairwise distinct Bethe roots, the number of singular and physical singular solutions. The expected number of physical solutions is of course known from simple representation theory of the symmetry algebra.} even for the simplest Heisenberg XXX spin chain. In order to find the number of solutions for fixed quantum numbers, one needs to solve BAE numerically and find all the solutions explicitly (see for example \cite{Hao:2013jqa}). From the algebraic geometry point of view, the number of solutions is nothing but the dimension of the quotient ring of BAE which will be defined in section\,\ref{sec:AG}. The quotient ring of BAE is a finite dimensional linear space whose dimension can be found without solving any equations ! We propose a method based on Gr\"obner basis to find the dimension of the quotient ring efficiently.\par

The second type of problem appears more recently in the context of integrability in AdS/CFT \cite{Vieira:2013wya,Basso:2017khq}. We will give a more detailed introduction to the background of this problem in section\,\ref{sec:sum_over_BAE} and appendix\,\ref{sec:OPE}. The problem can be formulated as the follows. Let us consider a set of BAE with fixed quantum numbers and some additional constraints\footnote{Such as the condition that the total momentum of the state should be zero.} on rapidities. Typically the number of physical solutions is not unique. Consider a rational function $F(u_1,\cdots,u_N)$ of the rapidities. The problem is to compute the sum of the function $F(u_1,\cdots,u_N)$ evaluated at all physical solutions. The usual way to proceed is first solving BAE numerically and then plugging the solutions in $F(u_1,\cdots,u_N)$ and finally performing the sum. We propose a different approach which avoids solving BAE. The main point is that the function $F(u_1,\cdots,u_N)$ evaluated at the solutions of BAE can be mapped to a finite dimensional matrix called the companion matrix in the quotient ring. The summation over all physical solutions corresponds to taking the trace of this matrix. Importantly, the companion matrix can be constructed in purely algebraic way.\par

We would also like to mention that similar computational
algebro-geometric methods for summing over solutions have been applied to a rather different
field, which is the scattering amplitudes \cite{Huang:2015yka,
  Sogaard:2015dba,Bosma:2016ttj}. In the framework of Cachazo-He-Yuan
formalism \cite{Cachazo:2013hca, Cachazo:2013gna, Cachazo:2013iea,
Cachazo:2014nsa, Cachazo:2014xea}, the scattering amplitudes can be written as a sum of a given function over all possible solutions of the scattering equations. The scattering equations are also a set of algebraic equations like BAE\footnote{In fact, the set of scattering equations is strikingly similar to the Bethe ansatz equations of Gaudin model.} which can be studied by algebraic geometry. Compared to our case, the scattering equations are much simpler and the structure of the solutions are easier to study. For example, the number of physical solutions can be determined readily and an analytic formula is known.\par

The rest of this paper is structured as follows. In section\,\ref{sec:AG}, we review some basic algebraic geometry that is necessary to understand our methods. In section\,\ref{sec:completeness}, we study the completeness problem of Bethe ansatz by algebro-geometric methods. We first give a detailed discussion of the physical problem and then provide the method to count the solutions of BAE under additional constraints. In section\,\ref{sec:sum_over_BAE}, we propose an analytical method to compute the sum of a given function evaluated at all physical solutions of BAE with fixed quantum numbers. We conclude in section\,\ref{sec:conlcude} and give a list of open problems and future directions. More backgrounds and technical details are presented in the appendices.

\section{Basics of algebraic geometry}
\label{sec:AG}

In this section, we briefly review some rudiments of algebraic
geometry. We refer to
\cite{MR0463157,MR1288523,MR3330490,opac-b1094391} for the
mathematical details.
See also the lecture notes \cite{Zhang:2016kfo} for the application of computational algebraic
geometry for polynomial reductions in scattering amplitudes.

\subsection{Polynomial ring, ideal and affine variety}
Consider a polynomial ring $A_K= K[z_1, \ldots z_n]$. An ideal
$I$ of $A$ is a subset of $A$ such that,
\begin{enumerate}
\item $f_1+f_2\in I$, if $f_1\in I$ and $f_2\in I$.
\item $g f \in I$, for $f \in I$ and $g \in A$.
\end{enumerate}

A polynomial ring is a {\it Noether} ring, which means any ideal $I$
of $A$ is finitely generated: for an ideal $I$, there exist a finite
number of
polynomials $f_i\in I$ such that any polynomial $F\in I$ can be
expressed as
\begin{equation}
  \label{eq:1}
  F= \sum g_i f_i, \quad g_i \in A.
\end{equation}
We may write $I=\langle f_1, \ldots , f_k \rangle$. Given an ideal
$I$, we define {\it quotient ring} $A/I$ as the quotient set
specified by the equivalence relation: $f\sim g$ if and only if $f -g\in
I$.

We are interested in the common solutions of polynomial equations, or in
algebraic geometry language, the {\it algebraic set}. The algebraic set
$\mathcal Z(S)$ of a
subset $S$ of $A$ is the set in affine space $\bar K^n$,
\begin{equation}
  \label{eq:2}
  \mathcal Z_{\bar K}(S) \equiv \{p\in \bar K^n| f(p)=0,\ \forall f \in S\}
\end{equation}
Here $\bar K$ is a field extension of the original field $K$, since
frequently we need a field extension to get all the solutions.

It is clear that the algebraic set of polynomials is the same as the
algebraic set of the
ideal generated by these polynomials,
\begin{equation}
  \label{eq:3}
  \mathcal Z_{\bar K} (S) =  \mathcal Z_{\bar K} (\langle S \rangle ) \,.
\end{equation}
Therefore, we usually only consider the algebraic set of an ideal.

\subsection{Gr\"obner basis and quotient ring}
An ideal $I$ of $A$ can be generated by different generating sets, or basis.
In many cases, a ``convenient'' basis is needed. For polynomial
equation solving and polynomial reduction problems, the convenient
basis is the so-called {\it Gr\"obner basis }. A Gr\"obner basis is
an analog of the row echelon form in linear algebra, because it makes
the reduction in a polynomial ring possible. (Schematically, the polynomial
reduction towards an arbitrary generating set is ill-defined since the
result is non-unique, while the polynomial reduction towards a
Gr\"obner basis provides the unique result.)

To define a Gr\"obner
basis, we first need to define monomial orders in a polynomial ring. A monomial orders $\prec$ is a total order for all monomials in $A$
such that,
\begin{itemize}
\item if $u\prec v$ then for any monomial $w$, $uw \prec vw$.
\item if $u$ is non-constant monomial, then $1 \prec u$.
\end{itemize}
Some common monomial orders are {\it lex} (Lexicographic), {\it deglex}
  (DegreeLexicographic), and  {\it degrevlex}
  (DegreeReversedLexicographic). Given a monomial order $\prec$, for
  any polynomial $f\in A$ there
  is a unique {\it leading term}, $\LT(f)$ which is the highest monomial of $f$
  in the order $\prec$.

A G\"obner basis $G(I)$ of an ideal $I$ with respect to a monomial
order $\prec$ is a generating set of $I$ such that for any $f\in I$,
\begin{equation}
  \label{GB}
  \exists g_i\in G(I),\quad \LT(g_i)|\LT(f) .
\end{equation}
(Here $a|b$ means a monomial $b$ is divisible by another monomial
$a$). Given a monomial order $\prec$, the corresponding  G\"obner basis can
be computed by the Buchberger algorithm
\cite{Buchberger:1976:TBR:1088216.1088219} or more recent F4/F5 \cite{FAUGERE199961,Faugere:2002:NEA:780506.780516}
algorithms. Furthermore, for an ideal $I$, give a monomial order $\prec$, the so-called
{\it minimal reduced} G\"obner basis is unique. We give more details on the computation of Gr\"obner basis in appendix\,\ref{sec:Gr}.

The property \eqref{GB} ensures that the {\it polynomial division } of
a polynomial $F\in A$ towards an ideal $I$ in the order $\prec$, is well-defined:
\begin{equation}
  \label{eq:5}
  F=\sum a_i g_i +r
\end{equation}
where $g_i$'s are the elements of the Gr\"obner basis. $r$ is called
the remainder, which contains monomials {\it not} divisible by any
$\LT(g_i)$. Given the monomial order $\prec$, the remainder $r$ for
$F$ is unique.

Therefore, the polynomial division and Gr\"obner basis method provide
the {\it canonical} representation of elements in the quotient ring $A/I$. For two
polynomials $F_1$ and $F_2$, $[F_1]=[F_2]$ in $A/I$ if and only if
their remainders of the polynomial division are the same,
$[r_1]=[r_2]$.  In particular, $f\in I$ if and only if its remainder
of the polynomial division is zero. This is a very useful application
of Gr\"obner basis since it efficiently determines if a polynomial is
inside the ideal or not.

\subsection{Zero dimensional ideal}
\label{sec:zero_dim_ideal}
A zero dimensional ideal is a special case of ideals such that its
algebraic set in an {\it algebraic closed field} is a finite set, i.e.,
$|\mathcal Z_{\bar K}(I)|<\infty$. The study of zero dimensional ideals are crucial
for our Bethe Ansatz computations.

One of the important properties of a zero dimensional ideal $I$ define over $K$ is that the
number of solutions (in an algebraically closed field) equals the linear
dimension of the quotient ring
\begin{equation}
|\mathcal Z_{\bar K}(I)| = \dim_K (A_K/I)
\end{equation}
Note that the field $K$ need not be
algebraically closed, but the field extension $\bar K$ must be algebraically
closed for this formula. Let $G(I)$ be the Gr\"obner basis of $I$ in
any monomial ordering. Since $(A_K/I)$ is linearly spanned
by monomials which are not divisible by any elements in $\LT(G(I))$, the
number of solutions, $|\mathcal Z_{\bar K}(I)|$ equals the number of
monomials which are not divisible by $\LT(G(I))$. This statement
provides a valuable method of determining the number of solutions. In
practice, we can use the lattice algorithm \cite{MR3330490} to list
these monomials. If we only need the dimension $\dim_K (A_K/I)$, we
can use the command 'syz' in \texttt{Singular} \cite{DGPS}.

Let $(m_1, \ldots, m_k)$ be the monomial basis of $A_K/I$ determined from the
above Gr\"obner basis $G(I)$. We can reformulate the algebraic structure
of  $(A_K/I)$ as matrix operations. For any $f\in A_k$,
\begin{equation}
  \label{eq:4}
   [f] [m_i] = \sum_{j=1}^k [m_j] c_{ji},\quad c_{j}\in K,\quad i=1,\ldots,k
\end{equation}
The $k\times k$ matrix $c_{ji}$ is called the {\it companion matrix}. We denote the companion matrix
of the polynomial $f$ by $M_f$. It is clear that $M_f$=$M_g$ if and only if $[f]=[g]$ in $A/I$ and
\begin{align}
\label{eq:MM}
M_{f+g}=M_f+M_g,\quad
M_{f g}=M_fM_g=M_gM_f,.
\end{align}
Furthermore, if a polynomial $f$ is in the ideal $\langle g\rangle +
I$, we say the fraction $f/g$ is a ``polynomial'' in the quotient ring
$A/I$ by the abuse of terminologies. The reason is that, in this case,
\begin{equation}
  \label{eq:12}
  f= g q+ s, \quad s\in I\,.
\end{equation}
Hence in the quotient ring $A/I$, $[f]=[g] [q]$. For a point $\xi
\in \mathcal Z(I)$, if $g(\xi)\not=0$, then $f(\xi)/g(\xi)=q(\xi)$. In
this sense, the computation of a fraction over the solution set is
converted to the computation of a polynomial over the solutions.

Furthermore, we define $M_{f/g}\equiv M_q$. It is clear that when $M_g$ is
an invertible matrix,
\begin{equation}
\label{eq:Mfbyg}
  M_{f/g}=M_f M_g^{-1}\,.
\end{equation}

Companion matrix is a powerful tool for computing the sum of values of
$f$ evaluated at the algebraic set (solutions) of $I$ over the
algebraically closed field extension $\bar K$. Let $(\xi_1,\ldots,
\xi_k)$ be the elements of $|\mathcal Z_{\bar K}(I)|$,
\begin{equation}
  \label{eq:6}
  \sum_{i=1}^k f(\xi_i) = \tr M_f
\end{equation}
Hence this sum over solutions over $\bar K$ can be evaluated directly
from the Gr\"obner basis over the field $K$. It also proves that this
sum must be inside $K$, even though individual terms may not be.


\section{Application I. Completeness of Bethe ansatz }
\label{sec:completeness}
As a first application of algebro-geometric approach, we revisit the completeness problem of Bethe ansatz in this section. The main calculation is to count the number of solutions of BAE under additional constraints. The usual way of finding the number of solutions is by solving the equations numerically and finding all the solutions explicitly \cite{Hao:2013rza,Hao:2013jqa}. However, if our aim is simply counting the number of solutions, this approach is overkilling. Using algebro-geometric approaches, we can avoid solving BAE and reduce the computation to simple algebraic manipulations.\par

We start by a detailed discussion on the completeness of Bethe ansatz, using the Heisenberg XXX spin chain as our example. Our goal is to explain why certain kinds of solutions of BAE are `non-physical' and should be discarded. After that, we present a methods based on Gr\"obner basis and the quotient ring to count the number of solutions.

\subsection{Completeness of Bethe ansatz for XXX spin chain}
Many integrable models can be solved by Bethe ansatz \cite{Bethe:Anstaz}. In practice this means one has a systematic method to construct the eigenstates of the Hamiltonian and compute the corresponding eigenvalues. The completeness problem of Bethe ansatz is whether all the eigenstates of the Hamiltonian can be constructed by Bethe ansatz. This question turns out to be quite subtle and there is no general answer to it.\par

In this subsection, we consider the completeness of Bethe ansatz for $SU(2)$ invariant Heisenberg XXX spin chain in the spin-$\frac{1}{2}$ representation. There has been arguments for the completeness of Bethe ansatz in the thermodynamic limit where the length of the spin chain is infinite \cite{Bethe:Anstaz,TakahashiComplete,kirillov1985combinatorial,Faddeev:ABA}. These arguments are based on the string hypothesis, which needs justification itself. The arguments lead to the correct number of states in the thermodynamic limit but were challenged in the more recent work \cite{Hao:2013jqa}, it is thus still unclear how to justify this kind of arguments in a more rigorous way. When the length of the spin chain is finite, the problem is more difficult and has been investigated in \cite{Avdeev1986,Essler,Sid} (see also \cite{Noh,Fabricius:2000yx,Baxter:2001sx,Tarasov}). In \cite{Hao:2013jqa} a conjecture for the number of solutions with pairwise distinct roots in terms of the number of singular solutions is proposed. This conjecture has been checked by solving BAE numerically up to $L=14$ (see \cite{Hao:2013rza} for a generalization to higher spin representations and \cite{Kirillov_rigged,Kirillov_remark} for relations with rigged configurations). We will review this conjecture below. Following this approach, the statement of completeness of Bethe ansatz can be formulated in terms of numbers of solutions of BAE with various additional constraints.\par

The Heisenberg XXX spin chain is a one-dimensional quantum lattice model with the following Hamiltonian
\begin{align}
\label{eq:Heisenberg_H}
H_{\text{XXX}}=\frac{1}{4}\sum_{j=1}^L(\vec{\sigma}_j\cdot\sigma_{j+1}-1),\qquad \vec{\sigma}_{L+1}=\vec{\sigma}_1
\end{align}
where $L$ is the length of the spin chain and we have imposed periodic boundary condition. Here $\vec{\sigma}=(\sigma_1,\sigma_2,\sigma_3)$ are the $2\times 2$ Pauli matrices and $\vec{\sigma}_k$ denotes the spin operator at position $k$. At each site, the spin can point either up or down, so the Hilbert space has dimension $2^L$. The Heisenberg spin chain can be solved by Bethe ansatz \cite{Bethe:Anstaz,Faddeev:ABA}. In this approach, each eigenstate is labeled by a set of variables $\{u_1,\cdots,u_N\}$ called the rapidities where $N$ is the number of flipped spins. The rapidities satisfy the following BAE
\begin{align}
\label{eq:BAEsu2}
\left(\frac{u_j+i/2}{u_j-i/2}\right)^L=\prod_{k\ne j}^N\frac{u_j-u_k+i}{u_j-u_k-i},\qquad j=1,\cdots,N.
\end{align}
The corresponding eigenvalue is given by
\begin{align}
\label{eq:eigenvalueH}
E_N=-\frac{1}{2}\sum_{k=1}^N\frac{1}{u_k^2+1/4}
\end{align}
Naively, one might expect that each solution of BAE corresponds to an eigenstate. However, this is not true and there are solutions of BAE which one should discard. In particular, the following four kinds of solutions need special care
\begin{enumerate}
\item \textbf{Coinciding rapidities}. The BAE allows solutions where two of the rapidities coincide, namely $u_i=u_j$ for some $u_i,u_j\in\{u_1,\cdots,u_N\}$. For Heisenberg spin chain (\ref{eq:Heisenberg_H}), these solutions are not physical and should be discarded. However, we want to mention that whether this kind of solutions are allowed or not in fact depends on the model under consideration \cite{Hao:2013rza}.
\item \textbf{Solutions with $N>L/2$}. The BAE (\ref{eq:BAEsu2}) can be solved for any $N\le L$. However, when we count the number of physical solutions, we do not consider the cases with magnon number $N>L/2$. This is because the eigenvectors corresponding to these solutions are not independent from the ones with $N\le L/2$.
\item \textbf{Solutions at infinity}. The BAE also allows solutions at infinity, namely we can take some $u_i\to\infty$. This case corresponds to the descendant states which are necessary for the completeness of Bethe ansatz. However, when we consider the solutions of BAE, we usually count the number of primary states, \textit{i.e.} no roots at infinity. The number of descendant states of a given primary state can be counted straightforwardly.
\item \textbf{Singular solutions}. There are also solutions of BAE at which the eigenvalues diverge (\ref{eq:eigenvalueH}) and the eigenstates are also singular. These solutions are called \emph{singular solutions}. To determine whether a singular solution is physical or not, one needs to perform a careful regularization. As it turns out, some of the singular solutions are physical and the others are not. The conditions for physical singular solutions are given in \cite{Nepomechie:2013mua}, which we quote in (\ref{eq:phys_sings}).
\end{enumerate}
For the readers' convenience, we give more detailed discussions on the above points in appendix\,\ref{app:BAEcompleteness}.\par


For the algebro-geometric approach, there's an additional subtlety which is the non-trivial multiplicities of certain solutions. While it is quite normal for algebraic equations to have solutions with multiplicities greater than one, physically we count them as one solution. The number of solutions is counted with multiplicity in algebro-geometric methods and we need to get rid of the multiplicities when counting the number of physical solutions.\par

By solving BAE for a few cases, we find that the multiple solutions are the ones contain $u_j=\pm i/2$, which are the singular solutions. In order to obtain the correct counting, our strategy is to consider separately the singular solutions and the rest ones. To obtain non-singular solutions, we introduce an auxiliary variable $w$ and add the constraint
\begin{align}
\label{eq:additionalCons}
w\prod_{j=1}^N(u_j^2+1/4)-1=0
\end{align}
to the original set of BAE. We see that whenever $u_j=\pm i/2$, (\ref{eq:additionalCons}) cannot be satisfied. To obtain the singular solutions, we put $u_1=i/2$ and $u_2=-i/2$ and solve for the rest variables.\par

Finally, the completeness of Bethe ansatz can be formulated as a statement of the numbers of solutions of BAE under various constraints. Let us denote the number of pairwise distinct (\emph{Pauli principle}) finite solutions (\emph{primary state}) for $N\le L/2$ by $\mathcal{N}_{L,N}$. Among these solutions, we denote the number of singular solutions by $\mathcal{N}_{L,N}^{\,\text{s}}$ and the singular physical solutions by $\mathcal{N}_{L,N}^{\,\text{sphy}}$. The number of solutions are counted \emph{without multiplicities}. The statement of completeness of Bethe ansatz is \cite{Hao:2013jqa}
\begin{align}
\label{eq:conjecture_numbers}
\mathcal{N}_{L,N}-\mathcal{N}_{L,N}^{\,\text{s}}
+\mathcal{N}_{L,N}^{\,\text{sphys}}=
{L\choose N}-{L\choose N-1}.
\end{align}
This is the alluded conjecture in \cite{Hao:2013jqa}. It has been confirmed by numerics up to $L=14$.\par

The goal of algebro-geometric approach is twofold. The first goal is to provide more efficient and stable methods to find the number of solutions $\mathcal{N}_{L,N}$, $\mathcal{N}_{L,N}^{\,\text{s}}$ and $\mathcal{N}_{L,N}^{\,\text{sphys}}$ for given $L$ and $N$ and test the conjectures further. The second and more ambitious goal is to find analytical expressions for these numbers in terms of $L$ and $N$. This requires a careful use of some powerful theorems in algebraic geometry such as the BKK theorem \cite{Bernshtein,Kushnirenko,Khovanskii}. While the second goal is not yet achieved in the current work and is still under investigation, we provide an efficient method for the first goal in what follows.

\subsection{Counting the number of solutions}
In this section, we explain how to apply the method of Gr\"obner basis to compute the numbers $\mathcal{N}_{L,N}$, $\mathcal{N}_{L,N}^{\,\text{s}}$ and $\mathcal{N}_{L,N}^{\,\text{sphys}}$ for given $L$ and $N$. The basic idea is that the number of solutions for a given set of polynomial equations is the dimension of the corresponding quotient ring. Instead of solving equations, we construct the quotient rings and compute their dimensions.\par

For a given $L$ and $N$, let us define the following polynomials.
\begin{align}
\rB_j=&\,(u_j+i/2)^LQ_{\mathbf{u}}(u_j-i)+(u_j-i/2)^LQ_{\mathbf{u}}(u_j+i),\qquad j=1,\cdots,N\\
\rB=&\,w\,(u_1^2+1/4)\cdots(u_N^2+1/4)-1\\
\rB'=&\,w\,(u_3^2+1/4)\cdots(u_N^2+1/4)-1
\end{align}
where $Q_{\mathbf{u}}(u)$ is the Baxter polynomial defined by
\begin{align}
Q_{\mathbf{u}}(u)=\prod_{k=1}^N(u-u_k).
\end{align}
To have pairwise distinct roots, we define the polynomials
\begin{align}
\rA_{ij}=\frac{\rB_i-\rB_j}{u_i-u_j},\qquad i=1,\cdots,N-1;\,\,j=i+1,\cdots,N.
\end{align}
This is a classical trick of getting distinct roots in algebraic
geometry. For singular and singular physical solutions, we define the following polynomials
\begin{align}
\rS_k=&\,\phantom{+}(u_k+i/2)^{L-1}(u_k-3i/2)\prod_{j=3}^N(u_k-u_j-i)\\\nonumber
&+(u_k-i/2)^{L-1}(u_k+3i/2)\prod_{j=3}^N(u_k-u_j+i),\qquad k=3,\cdots,N,\\\nonumber
\rS=&\,\prod_{k=3}^N(u_k+i/2)^L+(-1)^{L+1}\prod_{k=3}^N(u_k-i/2)^L.
\end{align}
Using these polynomials, we define the following ideals
\begin{align}
\label{eq:idealsSU2}
I_{\text{NS}}=&\,\langle \rB_1,\cdots,\rB_N,\rB,\rA_{12},\cdots,\rA_{N-1,N}\rangle,\\\nonumber
I_{\text{S}}=&\,\langle \rS_3,\cdots,\rS_N, \rB' ,A_{34},\cdots,\rA_{N-1,N}\rangle,\\\nonumber
I_{\text{SP}}=&\,\langle \rS_3,\cdots,\rS_N, \rB', \rS,A_{34},\cdots,\rA_{N-1,N}\rangle
\end{align}
where the subscribes denote `Non-Singular', `Singular' and `Singular Physical'. The corresponding quotient rings are defined as
\begin{align}
Q_{\text{NS}}=&\,\mathbb{C}[u_1,\cdots,u_N]/I_{\text{NS}},\\\nonumber
Q_{\text{S}}=&\,\mathbb{C}[u_3,\cdots,u_N]/I_{\text{S}},\\\nonumber
Q_{\text{SP}}=&\,\mathbb{C}[u_3,\cdots,u_N]/I_{\text{SP}}.
\end{align}
All the three quotient rings are finite dimensional linear spaces. The numbers $\mathcal{N}_{L,N}$, $\mathcal{N}_{L,N}^{\,\text{s}}$ and $\mathcal{N}_{L,N}^{\,\text{sphys}}$ are given in terms of the dimensions of the quotient rings as
\begin{align}
\mathcal{N}_{L,N}=\frac{\dim Q_{\text{NS}}}{N!}+\frac{\dim Q_{\text{S}}}{(N-2)!},\qquad
\mathcal{N}_{L,N}^{\,\text{s}}=\frac{\dim Q_{\text{S}}}{(N-2)!},\qquad
\mathcal{N}_{L,N}^{\,\text{sphys}}=\frac{\dim Q_{\text{SP}}}{(N-2)!}
\end{align}
We divide the dimensions by factorials to get rid of the permutation redundancy. Any permutation of the set of Bethe roots is considered to be the same solution, yet they correspond to different points in the affine variety. From the definitions of the ideals (\ref{eq:idealsSU2}), it is straightforward to compute the corresponding Gr\"obner basis. Then we can construct the standard basis for the quotient rings and the dimensions of the quotient rings follows.\par

\subsection{A symmetrization trick}
Note that BAE (for non-singular and
singular solutions) is totally symmetric
in $u_1, \ldots u_n$, i.e., the ideal for BAE is
symmetric under the full permutation group of $u_i$'s. We can take
advantage of this feature to speed up the Gr\"obner basis
computation. One immediate choice is to apply the symmetric ideal
Gr\"obner algorithm, ``symodstd.lib'' in Singular. However, this
approach is still not fast enough for our propose. Instead, we
discovered the following trick:

For a totally symmetric ideal $I$ in variables $u_1, \ldots u_n$, we
add $n$ auxiliary variables $s_1, \ldots s_n$ and $n$ auxiliary
equations to make a new ideal $\tilde I$,
\begin{equation}
  s_k-\sum_{j_1<\ldots j_k} u_{j_1}  \ldots u_{j_k} =0,\quad
  k=1,\ldots ,n\;.
\label{symmetrization}
\end{equation}
Therefore we define $s_k$ as the $k$-th elementary polynomials in $u_1, \ldots
u_n$. We find that with auxiliary variables and equations, and a block
order
$[u_1, \ldots
u_n]\succ [s_n, \ldots
s_1]$, the
Gr\"obner basis computation is much faster. Furthermore the
resulting Gr\"obner basis for $\tilde I$ is much shorter comparing
with that for $I$. We believe that the improvement comes from the
fact that BAE is much simpler in terms of the symmetric
variables  $s_1, \ldots, s_n$. The solutions of $\tilde I$' are in
one-to-one correspondence to the solution of $I$, so this method
is sufficient.

As a very interesting byproduct, this trick provides a new representation of BAE: The Gr\"obner basis $G(\tilde I)$, in the block
order mentioned above, eliminates the original variables $u_1 ,\ldots
u_n$ and gives a set of equations only in  $s_1, \ldots, s_n$.
\begin{equation}
  \label{eq:7}
  \langle G(\tilde I) \cap \mathbb K[s_1,\ldots s_n] \rangle = \tilde I \cap \mathbb K[s_1,\ldots s_n]\,.
\end{equation}
(On the left hand side of the equation, $\langle \ldots \rangle$ means the
ideal inside $K[s_1,\ldots s_n]$.) Usually the symmetrized BAE in $s_1, \ldots, s_n$ is
simpler than the original one since the permutation symmetry
group $S_n$ is removed. For instance, consider the $L=8, N=4$ BAE for nonsingular roots. This trick provides the new set of symmetrized BAE,
\begin{align}
  \label{eq:8}
 \mathcal S:\ 552960 s_4^3-76032 s_4^2+26496 s_2 s_4-8048 s_4+2400 s_2^2+21888 s_3^2-1848
   s_2-671&=0,\nonumber\\
432 s_2^2+4608 s_4 s_2-336
   s_2+3312 s_3^2+11520 s_4^2+20736 s_3^2 s_4-2208 s_4-119&=0,\nonumber
\\2304 s_3 s_4^2+576 s_2 s_3 s_4+12 s_2 s_3-s_3&=0,\  \nonumber
\\ 96 s_3^3+6 s_2 s_3+288 s_2 s_4
   s_3-16 s_4 s_3-3 s_3&=0,\nonumber\\-144 s_2^2+20736 s_4^2 s_2-1152 s_4 s_2+111 s_2-1152 s_3^2-4608
   s_4^2+528 s_4+41&=0,\nonumber\\-144 s_2^2+10368 s_3^2 s_2+12672 s_4 s_2+120 s_2-1152 s_3^2-11520
   s_4^2+1248 s_4+59&=0,\nonumber\\864 s_4 s_2^2-18 s_2^2-1008 s_4
   s_2+15 s_2-144 s_3^2+48 s_4+4&=0,\nonumber\\\ 3 s_3
   s_2^2-3 s_3 s_2-4 s_3 s_4&=0,\nonumber\\48 s_2^3-48 s_2^2-352 s_4 s_2-6 s_2-96 s_3^2+16
   s_4+3&=0,\nonumber\\\ s_1&=0 \,.
\end{align}
These equations have at most polynomial degree $3$ while the
original BAE has degree $10$. Furthermore,
$\mathcal S$ in the $s_1, \ldots, s_4$ coordinate has $11$
solutions, and correctly counts the number of nonsingular Bethe roots,
without permutation redundancy. On the other hand, the original BAE formally has $264$ solutions and we have to divide
this number by $4!$ to get the correct counting $11$ without permutations.

In most cases, physical quantities
are symmetric functions of $u_1, \ldots u_n$ and hence a
function in the elementary polynomials $s_1, \ldots s_n$, the above
new form of BAE in $s_1, \ldots s_n$ is sufficient
for physical purposes and makes computations much easier.

The Bethe roots counting results are given in the following table:
\begin{table}[h!]
  \centering
  \begin{tabular}{|c|c|c|c|c|}
  \hline
   $L$  & $N$ & $\mathcal{N}_{L,N}^{\,\text{ns}}$ & $\mathcal{N}_{L,N}^{\,\text{s}}$ & $\mathcal{N}_{L,N}^{\,\text{sphys}}$\\
\hline
 $6$ & $3$ &$9$ & $5$ & $1$ \\
\hline
$7$ & $3$ &$20$ & $6$ & $0$\\
\hline
$8$ & $3$ &$34$ & $7$ & $1$ \\
\hline
$8$ & $4$ &$32$ & $21$ & $3$\\
\hline
$9$ & $4$ &$69$ & $27$ & $0$ \\
\hline
$10$ & $5$ &$122$ & $84$ & $4$\\
\hline
$12$ & $5$ &$455$  & $163$ & $5$\\
\hline
$12$ & $6$ &$452$  & $330$ & $10$\\
\hline
  \end{tabular}
  \caption{Counting number of Bethe roots with Gr\"obner basis. Here $\mathcal{N}_{L,N}^{\,\text{ns}}$ denotes the number of nonsingular solutions for given $L$ and $N$.}
  \label{Counting}
\end{table}
these numbers agree with Table 2. of \cite{Nepomechie:2013mua} except
for the case $L=12$ and $N=5$. \footnote{In this case,
  the ref. \cite{Nepomechie:2013mua} claims that there are $454$ nonsingular
  solutions, $163$ singular solutions and $6$ physical singular
  solutions. However, we double checked that there should be $5$ physical
  singular solutions by explicitly applying 'Solve' in Mathematica.}

On a laptop with 16GB RAM and one processor Intel Core i7 without
parallelization, we can perform the calculation up to $L=12$,
$N=6$. We use both the software {\rm Singular} \cite{DGPS} and {\rm
  FGb} \cite{FGb} for this
computation.

We comment that this method is very efficient: for example, it only
takes about $124$ seconds to get the Gr\"obner basis for the BAE with $L=12$ and $N=6$, on the laptop mentioned above
with the software {\rm FGb}. Notice that the authors of \cite{Nepomechie:2013mua} used clusters to compute
these numbers while we are simply using laptops.

Finally, we would like to mention that in parallel with the Gr\"obner basis method, it
is also possible to count the number of Bethe root with the so-called
{\it resultant} method. The details of this direction are beyond the scope of this
paper and we sketch it in the appendix\,\ref{sec:resultant}.

\section{Application II. Sum over solutions of BAE}
\label{sec:sum_over_BAE}
In this section, we study another kind of problem in integrable systems using algebro-geometric methods. Oftentimes, one encounters the problem of computing the following sum
\begin{align}
\label{eq:sumFF}
F=\sum_{\text{sol}}\mathcal{F}(u_1,\cdots,u_N)
\end{align}
where the summation runs over \emph{all} physical solutions\footnote{For a solution to be physical, one usually needs to impose extra selection rules, as was discussed in the completeness problem of BAE. Sometimes, when the quantity under consideration has more symmetry, one can restrict to even smaller subsects of solutions.} of BAE with fixed quantum numbers. For XXX spin chain, the quantum numbers are the length of the spin chain $L$ and the number of particles $N$. Here $\mathcal{F}(u_1,\cdots,u_N)$ is a rational function of the rapidities and might also depend on other parameters. One example of such function is the (square of) OPE coefficient in the planar $\mathcal{N}=4$ SYM theory which we will discuss below.\par

The usual way to proceed is first finding all the physical solutions of BAE numerically to very high precisions, plugging into the function $\mathcal{F}(u_1,\cdots,u_N)$ and then computing the sum numerically. An interesting observation in \cite{Vieira:2013wya} is that although each solution of BAE is a complicated irrational numbers and so is the resulting $\mathcal{F}(u_1,\cdots,u_N)$, when one sums over all the solutions, the final result gives a simple \emph{rational number} ! This observation was made by carefully looking at the numerical patterns in the final result.\par

The numerical approach has certain disadvantages. To start with, finding all solutions of BAE is a highly non-trivial task even for simple models. Secondly, due to numerical instabilities, it is not always easy to estimate to which precision should one be working with in order to find the pattern of rational numbers mentioned above. Finally, it is not clear whether the final result should be a rational number or not.\par

We propose an alternative method based on algebraic geometry to perform the sum (\ref{eq:sumFF}). Using this approach, there's no need to solve BAE and the computation is reduced to taking traces of numerical matrices whose matrix elements are rational numbers if the coefficients of $\mathcal{F}(u_1,\cdots,u_N)$ are rational numbers\footnote{Our notion of rational numbers also includes complex numbers whose real and imaginary parts are both rational.}, which is the case for OPE coefficients. It is then obvious that the final result should be a rational number.\par

In what follows, we first describe the general method with the help of a simple toy problem. Then we demonstrate how our method works in the context of \cite{Vieira:2013wya} and how to generalize it to higher loop orders in this case.

\subsection{Description of the method}
In this section, we present more details for the brief discussions on companion matrix in section\,\ref{sec:zero_dim_ideal} for our current problem.
We start with a set of polynomial equations (which we can think as BAE and additional constraints)
\begin{align}
\label{eq:BAEf}
\rF_1(u_1,\cdots,u_N)=\cdots=\rF_n(u_1,\cdots,u_N)=0
\end{align}
Let us denote the ideal generated by $\rF_1,\cdots,\rF_n$ by $I_{\rF}=\langle \rF_1,\cdots,\rF_n\rangle$. The corresponding quotient ring is defined by
\begin{align}
Q_\rF=\mathbb{C}[u_1,\cdots,u_N]/\langle \rF_1,\cdots,\rF_n\rangle.
\end{align}
Because the quotient ring $Q_\rF$ is a \emph{finite dimensional} linear space, it can be spanned by a set basis $\{e_1,\cdots,e_s\}$. Consider any polynomial $\mathcal{P}(u_1,\cdots,u_N)$ in $\mathbb{C}[u_1,\cdots,u_N]$. After imposing the `on-shell conditions' (\ref{eq:BAEf}), $\mathcal{P}(u_1,\cdots,u_N)$ becomes a function in the quotient ring $Q_\rF$ and can be represented in terms of a matrix called the \emph{companion matrix}. To be more precise, we have
\begin{align}
\mathcal{P}\cdot[e_i]=\sum_{j=1}^s\left(M_{\mathcal{P}}\right)_{ij}[e_j]
\end{align}
where $M_{\mathcal{P}}$ is a \emph{numerical matrix} of dimension $s\times s$. Here $[e_j]$ denotes the conjugacy class of the basis $e_j$ under the identification
\begin{align}
e_j\sim e_j+k,\qquad k\in I_\rF.
\end{align}
Our method is based on the following crucial result
\begin{align}
\label{eq:Fsum}
P=\sum_{\text{sol}}\mathcal{P}(u_1,\cdots,u_N)=\tr M_{\mathcal{P}}.
\end{align}
Two comments are in order. Firstly, the companion matrix $M_{\mathcal{P}}$ contains all the information about the on-shell quantity $\mathcal{P}(u_1,\cdots,u_N)$. If one diagonalizes the $s\times s$ matrix $M_{\mathcal{P}}$, each eigenvalue correspond to $\mathcal{P}(u_1,\cdots,u_N)$ with $u_1,\cdots,u_N$ at one of the physical solutions of (\ref{eq:BAEf}). Secondly, if the equations (\ref{eq:BAEf}) are symmetric with respect to some of the variables, we should divide the number by proper symmetric factors when performing the sum (\ref{eq:Fsum}) to get rid of permutation redundancy. An alternative way to get ride of the permutation redundancy is to rewrite the polynomial $\mathcal{P}(u_1,\cdots,u_N)$ in terms of elementary symmetric polynomials $\mathcal{P}^{\rs}(s_1,\cdots,s_N)$ and perform the calculation in the quotient ring of symmetrized BAE.\par

The main task is to construct the basis $\{e_j\}$ for $Q_\rF$ and find $M_{\mathcal{P}}$. This can be done using the Gr\"obner basis. Let us denote the Gr\"obner basis of $I_\rF$ to be $\rG_1,\cdots,\rG_n$, which can be computed from $\rF_1,\cdots,\rF_n$ and
\begin{align}
I_\rF=\langle \rF_1,\cdots,\rF_n\rangle=\langle \rG_1,\cdots,\rG_n\rangle.
\end{align}
Then the standard basis of quotient ring can be constructed by the method given in section\,\ref{sec:zero_dim_ideal}. The companion matrix can be constructed as follows. First multiply the polynomial $\mathcal{P}(u_1,\cdots,u_N)$ with one of the basis $e_j$ and then divide the result by the Gr\"obner basis,
\begin{align}
\mathcal{P}\cdot e_j=\sum_{k=1}^n a_k\,\rG_k+\mathcal{P}_j
\end{align}
where $a_n$ are polynomials in $u_1,\cdots,u_N$ and $\mathcal{P}_j$ is the remainder. Since $\{\rG_1,\cdots,\rG_n\}$ are Gr\"obner basis, the remainder $\mathcal{P}_j$ is well-defined. Now that $\mathcal{P}_j$ is a polynomial defined in the quotient ring $Q_{\rF}$, it can be expanded in terms of the standard basis as
\begin{align}
\mathcal{P}_j=\sum_{k=1}^s \left(M_{\mathcal{P}}\right)_{jk}\,e_k.
\end{align}
This gives the $j$-th row of the matrix $M_{\mathcal{P}}$. Repeating this process for $j=1,\cdots,s$, we obtain the matrix $M_{\mathcal{P}}$. In this way, we can construct the companion matrix of any polynomials. To find the companion matrices for rational functions, we can make use the properties of the companion matrices (\ref{eq:MM}) and (\ref{eq:Mfbyg}).

\subsection{A simple example}
To illustrate our approach, we consider a simple example. We first solve the problem by a numerical approach and then by our algebro-geometric approach in order to make a comparison.
Let us take
\begin{align}
\rF_1=x^4y^2+3xy+1,\qquad \rF_2=y^3+y^2-2
\end{align}
and
\begin{align}
\label{eq:F_example}
\mathcal{P}(x,y)=\frac{x^3}{3}+\frac{y^3}{7}+4xy(x+y)+2x+1.
\end{align}
It is easy to find that the equations $\rF_1=\rF_2=0$ have 12 solutions. Setting the working precision to 22 digits, we can find the numerical solutions quite easily
\begin{align}
x_1=&\,-0.9692939705422372032999 - 0.8607793416347397527029 i,\\\nonumber
y_1=&\,-1.000000000000000000000 - 1.000000000000000000000 i\\\nonumber
\qquad\vdots\\\nonumber
x_{12}=&\,0.822576433302391503774 + 1.260317961087082767027 i,\\\nonumber
y_{12}=&\,1.000000000000000000000.
\end{align}
Plugging into (\ref{eq:F_example}), we obtain 12 numerical values
\begin{align}
\label{eq:Fexample_Num}
\mathcal{P}(x_1,y_1)=&\,12.52841718172878750443 - 17.82690255958560159754 i,\\\nonumber
\qquad\vdots\\\nonumber
\mathcal{P}(x_{12},y_{12})=&\,1.31018389310726616255 + 16.04104511847801597154 i.
\end{align}
Finally we take the sum of the 12 values in (\ref{eq:Fexample_Num}) and obtain
\begin{align}
P=\sum_{i=1}^{12}\mathcal{P}(x_i,y_i)=14.\textcolor{blue}{\textbf{857142}}\textbf{857142}\textcolor{blue}{\textbf{857142}}9\cdots
\end{align}
where we see a clear pattern. After rationalization, we obtain simply
\begin{align}
F=\frac{104}{7}
\end{align}
which is indeed a simple rational number. Now we can do the computation using our approach. The Gr\"obner basis in this case can be computed by the built-in function of \texttt{Mathematica}
\begin{align}
\texttt{GroebnerBasis[\{F$_1$,F$_2$\},\{x,y\}]}
\end{align}
The result is the following Gr\"obner basis
\begin{align}
\rG_1=3xy^2+3xy+y+2x^4+1,\qquad \rG_2=y^3+y^2-2.
\end{align}
Now we can construct the standard basis for the quotient ring $Q_{\text{eg}}=\mathbb{C}[x,y]/\langle \rG_1,\rG_2\rangle$. Using the lexicographical ordering for monomials $x\succ y$, the standard basis of $Q_{\text{eg}}$ is given by
\begin{align}
\label{eq:standard_basis}
e_1=&\,x^3 y^2,& e_2=&\,x^3 y,& e_3=&\,x^3,&\\\nonumber
e_4=&\,x^2 y^2,& e_5=&\,x^2 y,& e_6=&\,x^2,&\\\nonumber
e_7=&\,x^1 y^2,& e_8=&\,x^1y,&  e_9=&\,x^1,&\\\nonumber
e_{10}=&\,y^2,& e_{11}=&\,y,& e_{12}=&\,1.&
\end{align}
Notice that the dimension of $Q_{\text{eg}}$ equals the number of solutions of $\rF_1=\rF_2=0$. The next step is to construct the companion matrix $M_{\mathcal{P}}$. Let us first consider $e_1$. It is straightforward to calculate that\footnote{For example, one can use built-in function \texttt{PolynomialReduce} in \texttt{Mathematica}.}
\begin{align}
\mathcal{P}(x,y)e_1=a_1 \rG_1+a_2\rG_2+\mathcal{P}_1
\end{align}
where
\begin{align}
\mathcal{P}_1=\frac{8}{7}x^3y^2-12 x^2 y^2+12 xy^2-4y^2-\frac{9}{7}x^3y-10xy+\frac{2}{7}x^3-\frac{1}{3}x^2-24x-2.
\end{align}
It can be expanded in terms of the basis (\ref{eq:standard_basis}) as
\begin{align}
\mathcal{P}_1=\sum_{j=1}^{12}\left(M_{\mathcal{F}}\right)_{1j}e_j
\end{align}
where
\begin{align}
\left(M_{\mathcal{P}}\right)_{1j}=\left(\frac{8}{7}, -\frac{9}{7}, \frac{2}{7}, -12, 0, -\frac{1}{3}, 12, -10, -24, -4, 0, -2\right)
\end{align}
Working out the other rows $\left(M_{\mathcal{P}}\right)_{ij}$ in the same way, we obtain
\begin{align}
M_{\mathcal{F}}=\tiny{\frac{1}{42}
\left(
\begin{array}{cccccccccccc}
 48 & -54 & 12 & -504 & 0 & -14 & 504 & -420 & -1008 & -168 & 0 & -84 \\
 6 & 54 & -54 & -7 & -511 & 0 & -504 & 0 & -420 & -42 & -210 & 0 \\
 -27 & -21 & 54 & 0 & -7 & -511 & -210 & -714 & 0 & 0 & -42 & -210 \\
 252 & 336 & -336 & 48 & -54 & 12 & -504 & 0 & -14 & 0 & -168 & 0 \\
 -168 & 84 & 336 & 6 & 54 & -54 & -7 & -511 & 0 & 0 & 0 & -168 \\
 168 & 0 & 84 & -27 & -21 & 54 & 0 & -7 & -511 & -84 & -84 & 0 \\
 -168 & 0 & 336 & 252 & 336 & -336 & 48 & -54 & 12 & 0 & 0 & -14 \\
 168 & 0 & 0 & -168 & 84 & 336 & 6 & 54 & -54 & -7 & -7 & 0 \\
 0 & 168 & 0 & 168 & 0 & 84 & -27 & -21 & 54 & 0 & -7 & -7 \\
 14 & 0 & 0 & -168 & 0 & 336 & 252 & 336 & -336 & 48 & -12 & 12 \\
 0 & 14 & 0 & 168 & 0 & 0 & -168 & 84 & 336 & 6 & 54 & -12 \\
 0 & 0 & 14 & 0 & 168 & 0 & 168 & 0 & 84 & -6 & 0 & 54 \\
\end{array}
\right)}
\end{align}
It is easy to verify that
\begin{align}
F=\tr M_{\mathcal{P}}=\frac{104}{7}.
\end{align}
We notice immediately that from the second approach, we directly
manipulate the polynomials in a purely algebraic way and there is no
need to solve any equations. Therefore we completely avoid all the
subtleties of numerical approach. As a bonus, it is clear that
the final result should be a rational number since all the
manipulations, including the computation of Gr\"obner basis and companion matrix, involve only simple addition, substraction, multiplication and division of rational numbers and there is no room to create irrational numbers from these operations.

\subsection{Sum rule of OPE coefficients}
\label{sec:sum-rule-OPE}
In this section, we revisit the calculation of \cite{Vieira:2013wya} for OPE coefficients in planar $\mathcal{N}=4$ Super-Yang-Mills theory ($\mathcal{N}=4$ SYM) using algebro-geometric approach. Let us first give the minimal background of this calculation. It is now well accepted that $\mathcal{N}=4$ SYM theory is integrable in the planar limit \cite{BigReview}. In practice this means one can use integrability-based methods to compute physically interesting quantities of the theory. For a conformal field theory like $\mathcal{N}=4$ SYM theory, the most fundamental quantities of interest are the so-called conformal data which consists of the scaling dimensions of all primary operators and the OPE coefficients among these operators.\par

In order to check the predictions of integrability-based methods, one needs to compare with results from other approaches, such as direct field theoretical calculations based on Feynmann diagrams. The most convenient source of data for OPE coefficients are the four-point functions of BPS operators, which are known up to three loops in perturbation theory (see \cite{Chicherin:2015edu} and references therein). By performing operator product expansions of the four-point functions, one has access to the information of OPE coefficients. However, it is usually hard to extract a single OPE coefficient from four-point functions. The best one can do is to give predictions for the so-called \emph{sum rules} defined in (\ref{eq:sum_rule}). We give more details of the OPE coefficients and sum rules in appendix\,\ref{sec:OPE}. To summarize, one needs to compute the following quantity
\begin{align}
\label{eq:sum_rule}
F_{\text{S}}=\sum_{\text{sol.}}\left(C_{\mathbf{u}}^{\bullet\circ\circ}\right)^2 e^{\gamma_{\mathbf{u}}}
\end{align}
where $C^{\bullet\circ\circ}_{\mathbf{u}}$ is the OPE coefficient of two BPS operators and one non-BPS operator and $\gamma_{\mathbf{u}}$ is the anomalous dimension of the non-BPS operator. They are both functions of the rapidities $\mathbf{u}\equiv\{u_1,\cdots,u_S\}$. The structure constant is given by
\begin{align}
\label{eq:OPEcoe_explicit}
C_{\mathbf{u}}^{\bullet\circ\circ}=\sqrt{\frac{L(l+N)(L-l+N)}{\rC_{l+N}^N\rC_{L-l+N}^N}}\left(1-\frac{\gamma_{\mathbf{u}}}{2}\right)\frac{\mathcal{A}_l}{\mathcal{B}},
\qquad \rC_{M}^N=\frac{M!}{N!(M-N)!}
\end{align}
where
\begin{align}
\label{eq:OPEcoe_A}
\mathcal{A}_l=\frac{1}{\sqrt{\prod_{j\ne k}^Sf(u_j,u_k)}\prod_{j=1}^S(e^{-ip(u_j)}-1)}\sum_{\alpha\cup\bar{\alpha}=\mathbf{u}}(-1)^{|\alpha|}
\prod_{u_j\in\bar{\alpha}}e^{-ip(u_j)}\prod_{u_j\in\alpha\atop u_k\in\bar{\alpha}}f(u_j,u_k)
\end{align}
and
\begin{align}
\label{eq:OPEcoe_B}
\mathcal{B}^2=\frac{1}{\prod_{j=1}^S\frac{\partial p(u_j)}{\partial u_j}}\det\left(\frac{\partial}{\partial u_j}\left[Lp(u_k)-i\sum_{l\ne k}^S\log S(u_k,u_l)  \right]  \right).
\end{align}
The quantities in the above expressions such as the momentum $p(u)$, the $S$-matrix $S(u,v)$ and $f(u,v)$ are known functions of the coupling constant $g$, where
\begin{align}
g^2=\frac{g_{\text{YM}}^2N_c}{16\pi^2}.
\end{align}
We expand these quantities at weak coupling when $g\to 0$ and consider the result up to 1-loop, namely $\mathcal{O}(g^2)$ order. We consider the leading order in this subsection and discuss the one-loop result in the next subsection. At the leading order, the various quantities are given by \begin{align}
e^{ip(u)}=\frac{u+i/2}{u-i/2},\qquad f(u,v)=\frac{u-v+i}{u-v},\qquad S(u,v)=\frac{u-v+i}{u-v-i}
\end{align}
The anomalous dimension $\gamma_{\mathbf{u}}$ only starts to contribute at one-loop order and is given by
\begin{align}
\gamma_{\mathbf{u}}=g^2\sum_{j=1}^S\frac{1}{u_j^2+1/4}.
\end{align}
Let us now consider the sum rule in (\ref{eq:sum_rule}). The OPE coefficients depend on four integers $L,S,l,N$ and a set of rapidities $\{u_1,\cdots,u_S\}$. For fixed $L$ and $S$, these rapidities satisfy the BAE of $SL(2)$ spin chain
\begin{align}
\label{eq:BAE_SL2}
\left(\frac{u_j+i/2}{u_j-i/2}\right)^L=\prod_{k\ne j}^S\frac{u_j-u_k-i}{u_j-u_k+i},\qquad j=1,2,\cdots,S.
\end{align}
In addition, we also need to impose the zero momentum condition
\begin{align}
\label{eq:zero_momenta}
\prod_{j=1}^S e^{ip(u_j)}=\prod_{j=1}^S\frac{u_j+i/2}{u_j-i/2}=1.
\end{align}
The summation in (\ref{eq:sum_rule}) runs over all possible solutions of (\ref{eq:BAE_SL2}) and (\ref{eq:zero_momenta}) for fixed $L$ and $S$. For generic values of $L$ and $S$, the solutions of (\ref{eq:BAE_SL2}) and (\ref{eq:zero_momenta}) are not unique. This is precisely the same type of problem which we discussed in the previous subsection. We can apply our method to perform this sum. Since the coefficients that appear in the sum rule are all rational numbers, it is guaranteed from our approach that the final result will be a rational number as well. We give more details on the implementation of our method in what follows.\par

We first write down a basis that generate the ideal $I_{\text{S}}$ corresponding to (\ref{eq:BAE_SL2}) and (\ref{eq:zero_momenta}). In order to obtain a polynomial basis, we can write BAE as $\rF_1=\cdots=\rF_S=0$ where
\begin{align}
\rF_j=(u_j+i/2)^L Q_{\mathbf{u}}(u_j+i)+(u_j-i/2)^L Q_{\mathbf{u}}(u_j-i),\qquad j=1,\cdots,S.
\end{align}
$Q_{\mathbf{u}}(u)$ is the Baxter polynomial. The zero momentum condition is equivalent to $\rF=0$ where
\begin{align}
\rF=\prod_{j=1}^S(u_j+i/2)-\prod_{j=1}^S(u_j-i/2).
\end{align}
Solving these constraints naively, there are solutions with coinciding roots. These solutions are not allowed since they are not physical. To eliminate these solutions, we need to impose extra constraints. These constraints can be imposed in various ways. For example, we can define the following polynomials
\begin{align}
\rK_{ij}=\frac{\rF_i-\rF_j}{u_i-u_j},\qquad i=1,\cdots,S-1;\,j=i+1,\cdots,S
\end{align}
and impose $\rK_{ij}=0$. The ideal $I_\rS$ is then given by
\begin{align}
I_\rS=\langle \rF_1,\cdots,\rF_S,\rF,\rK_{12},\cdots,\rK_{S-1,S}\rangle
\end{align}
The computations of the Gr\"obner basis of $I_{\rS}$ and the basis of the quotient ring $Q_{\rS}=\mathbb{C}[u_1,\cdots,u_S]/I_{\rS}$ are standard. Once the basis for the quotient ring has been constructed, we can follow the same method described in the previous subsection to construct the companion matrix for the summand
\begin{align}
\mathcal{F}(u_1,\cdots,u_S)=(C^{\bullet\circ\circ}_{\mathbf{u}})^2 e^{\gamma_{\mathbf{u}}}.
\end{align}
As an example, we can consider the case with $L=4,S=4,l=2,N=1$. In
this case there are 5 allowed solutions and the sum rule
(\ref{eq:sum_rule}) at the leading order is $F=16/63$. We find the
dimension of the quotient ring is $\dim Q_\rS=120=5\times 4!$. We use
the lattice algorithm \cite{MR3330490}, implemented in our {\rm
  Mathematica} code to determine the $120$ monomials in the basis of $Q_\rS$.
As we explained before, the $S!$ permutation redundancy is due to the fact that the BAE and zero momentum condition are completely symmetric with respect to all the rapidities. For this example, our method leads to a matrix $M_{\mathcal{F}}$ of $120\times 120$ which we will not write down explicitly.\par

The function $\mathcal{F}$ is a rational function and can be written as the ratio of two polynomials $\mathcal{F}=\mathcal{P}/\mathcal{Q}$. Let us denote their corresponding multiplication matrices as $M_{\mathcal{P}}$ and $M_{\mathcal{Q}}$. We then have $M_{\mathcal{F}}=M_{\mathcal{P}}\cdot M_{\mathcal{Q}}^{-1}$.

Taking the trace of the matrix, we confirm that
\begin{align}
F=\frac{1}{4!}\tr \left(M_{\mathcal{P}}\cdot M_{\mathcal{Q}}^{-1}\right)=\frac{16}{63}.
\end{align}
We checked several other examples and in all the cases, we reproduce
the same results as in \cite{Vieira:2013wya}.\par

To improve the efficiency, we can also use the symmetrization trick in
\eqref{symmetrization}. Define $s_i$ to be the $i$-th elementary symmetric
polynomials in $u_1,\ldots, u_4$, $i=1,\ldots 4$. After calculating
the Gr\"obner basis in the block ordering $[u_1,u_2,u_3,u_4]\succ [s_4,s_3,s_2,s_1]$, the new form of the
BAE is
\begin{align}
  \label{eq:9}
4 s_3-s_1&=0,\nonumber \\
5 s_1 s_2-14 s_1&=0, \nonumber \\
 80 s_1 s_4-3 s_1&=0,\nonumber \\
-3 s_1^2+s_2+144 s_2 s_4+320 s_4-1&=0, \nonumber \\
108 s_1^2+16128 s_4^2-232s_2-10752 s_4-11&=0,\nonumber \\
 -102 s_1^2+72 s_2^2+140s_2-112 s_4+31&=0, \nonumber \\
25 s_1^3-241 s_1&=0\,.
\end{align}
This symmetrized equation system only contains $5$ solutions and hence the $4!$
permutation redundancy is removed.

The structure constant is a rational function in $s_i$'s, since
it is symmetric in $u_i$'s. On the solutions, the structure constant
is reduced to
\begin{align}
 F&\to \frac{\mathcal P_s}{\mathcal Q_s}\nonumber \\
&=\frac{10 \left(6062953559631 s_1^2-12892045110000 s_2-583954414840000 s_4-653431597500\right)}{63 \left(1020845747568 s_1^2-42110437500 s_2-386898261950000 s_4+1685904440625\right)}\,.
\end{align}
We can calculate the companion matrices $M_{\mathcal P_s}$ and
$M_{\mathcal Q_s}$ in the variables $s_1, s_2, s_3, s_4$. Note that $M_{\mathcal P_s}$ and
$M_{\mathcal Q_s}$ are much simpler than  $M_{\mathcal P}$ and
$M_{\mathcal Q}$, since they are $5\times 5$ matrices instead of $120
\times 120$ matrices. Summing over all Bethe solutions, we get the
same result,
\begin{equation}
\label{eq:11}
\text{Tr}\,(M_{\mathcal{P}_s} M_{\mathcal{Q}_s}^{-1}) =\frac{16}{63}.
\end{equation}
Here the factor $4!$ is no longer needed.

Finally, let us comment on the efficiency of our method. For the current calculation, we focus on the $SL(2)$ sector. The BAE of $SL(2)$ spin chain is actually quite easy to solve numerically and the solutions have very nice behaviors, such as all the roots are real and there are no solutions with higher multiplicities.\par

The analytical counterpart of solving BAE and additional constraints is the construction of the quotient ring. In this specific case, numerical solution is actually faster than constructing the quotient ring. One of the main reasons for this is the permutation redundancy which grows factorially with the number of magnons. This difficulty can be overcome partially by the symmetrization trick described before, but we suspect that further improvements should be possible. Since the elementary symmetric polynomials of rapidities are nothing but the coefficients of Baxter polynomials. It is thus very natural to work with $Q$-systems instead of BAE. Very recently, Marboe and Volin \cite{Marboe:2016yyn,Marboe:2017dmb} proposed an efficient method to find Bethe roots based on the $QQ$-relations. Instead of solving BAE, they solve a system of zero remainder conditions (ZRC) where the unknowns are exactly the coefficients of Baxter polynomials. This approach also has other merits such as it automatically selects physical solutions. The method turns out to be quite efficient and works for a large family of spin chains.\par

Combining our approach and the methods in \cite{Marboe:2016yyn,Marboe:2017dmb}, we can in fact construct the quotient ring for ZRC of the corresponding $Q$-system. In this case, constructing the quotient ring is even faster than solving the ZRC ! For instance, we can construct the quotient ring of ZRC of $SU(2)$ $Q$-system for $L=14$, $N=7$ spin chain within several minutes on a laptop. The systematic study of the Gr\"obner basis and the corresponding quotient ring of rational $Q$-systems and their applications will be presented in an upcoming publication \cite{Jiang:AG_BAE2}.\par

After solving the BAE numerically or constructing the quotient ring analytically, one still need to compute the sum over all allowed solutions. This second step is basically trivial for numerical calculations but can be non-trivial for analytical approaches. The main reason is that the quantity we are dealing with, namely the sum rule is a complicated function of rapidities and the complexity grows exponentially with the number of magnon. Unless a simpler form of this quantity is given, this complexity is inherent to any analytical methods and should not be considered as the disadvantage of our algebro-geometric approach. One way to improve the efficiency is to decompose the quantity into smaller and simpler parts, compute the multiplication matrices for the smaller parts and then combine them together. The last step involves only manipulations of numerical matrices and can be done efficiently.

\subsection{Higher loops}
In this section, we discuss how to compute the sum rules at one-loop order. Consider the following sum
\begin{align}
\label{eq:sumrule_lambda}
F(\lambda)=\sum_{\text{sol.}}\mathcal{F}(u_1,\cdots,u_N;\lambda)
\end{align}
where the summation runs over the solution of a set of BAE and possibly other additional selection rules that depend on an extra parameter $\lambda$. The prototype of this kind of equations we have in mind are the Beisert-Staudacher asymptotic BAE \cite{Beisert:2005fw} and the zero momentum condition in $\mathcal{N}=4$ SYM where $\lambda$ plays the rule of the 't Hooft coupling constant. For simplicity, we assume that the function $\mathcal{F}(u_1,\cdots,u_N;\lambda)$ are rational functions in rapidities $\{u_j\}$. It depends on $\lambda$ explicitly as well as implicitly through the rapidities.\par

We consider the weak coupling limit $\lambda\to 0$ and develop a perturbative approach to compute the sum (\ref{eq:sumrule_lambda}). The leading order is considered to be solved by our approach presented before and we use this knowledge to solve higher loop orders perturbatively. Let us first focus on one-loop order. We assume the summand allows a perturbative expansion in $\lambda$
\begin{align}
\mathcal{F}(x_1,\cdots,x_N;\lambda)=\sum_{k=0}^\infty\mathcal{F}_k(x_1,\cdots,x_N)\lambda^k
\end{align}
At one-loop, the contribution comes from $\mathcal{F}_0$ and $\mathcal{F}_1$. We also perform a perturbative expansion of the Bethe roots
\begin{align}
u_j(\lambda)=u_j^{(0)}+\lambda u_j^{(1)}+\cdots
\end{align}
Finally the sum $F(\lambda)$ can also be expanded in $\lambda$
\begin{align}
F(\lambda)=\sum_{k=0}^\infty F_k\,\lambda^k
\end{align}
The leading order $F_0$ is considered to be known and are interested in $F_1$, which is simply given by
\begin{align}
\label{eq:one-loopF1}
F_1=\sum_{\text{sol }\{u_j^{(0)}\}}\left(\sum_{k=1}^N u_k^{(1)}\left.\frac{\partial}{\partial x_k}\mathcal{F}_0(x_1,\cdots,x_N)\right|_{x_k=u_k^{(0)}}  +\mathcal{F}_1(u_1^{(0)},\cdots,u_N^{(0)})\right)
\end{align}
where the sum is over \emph{leading order} solutions. This implies that we only need the quotient ring of the leading order, which is known. The second term in (\ref{eq:one-loopF1}) is explicit and are rational functions of $\{u_j^{(0)}\}$, which can be handled straightforwardly as before. The first term involves $u_j^{(1)}$ and we need to express them in terms of $u_k^{(0)}$. This can be achieved as follows. Consider the following BAE
\begin{align}
e^{ip(u_j)L}\prod_{k=1\atop k\ne j}^N S(u_j,u_k)=1
\end{align}
where both $p(u)$ and $S(u,v)$ depend on $\lambda$. We can expand the above BAE in $\lambda$ and obtain an approximated BAE valid at one-loop. Then we plug in the ansatz $u_j=u_j^{(0)}+\lambda u_j^{(1)}$ into the approximated BAE and expand up to one-loop order. The leading order BAE involves only $u_j^{(0)}$ and is considered to be solved. The one-loop BAE involves both $u_j^{(0)}$ and $u_k^{(1)}$. The important point is that the dependence of one-loop BAE in $u_k^{(1)}$ is \emph{linear}. Therefore we can regard $u_j^{(0)}$ as constants and solve the linear problem for $u_k^{(1)}$. This can be done straightforwardly and we obtain $u_j^{(1)}=U_j(u_1^{(0)},\cdots,u_N^{(0)})$ which in general is a rational function in $u_k^{(0)}$. After we plug $U_j$ back into (\ref{eq:one-loopF1}), the resulting expression is a rational function depending only on $u_j^{(0)}$ and the sum can be performed as before.\par

Generalization to higher loops orders is straightforward. The explicit part is easy to deal with. The main complexity of the implicit part comes from expressing $u_j^{(n)}$ in terms of $u_j^{(0)}$. This can be done in a recursive way. We first solve the approximated BAE at one-loop, finding $u_j^{(1)}$ in terms of $u_k^{(0)}$. Then we solve the approximated BAE at two-loop order using the ansatz $u_j(\lambda)=u_j^{(0)}+\lambda\,u_j^{(1)}+\lambda^2 u_j^{(2)}$. The approximated BAE at two-loop order involve $u_j^{(0)}$ and $u_j^{(1)}$ and depend on $u_j^{(2)}$ linearly. We can again solve the linear problem to find $u_j^{(2)}$ in terms of $u_k^{(1)}$ and $u_k^{(0)}$. Since we already know how to express $u_k^{(1)}$ in terms of $u_k^{(0)}$ from the previous order, we can express $u_j^{(2)}$ in terms of $u_k^{(0)}$. Therefore, to find expressions of $u_j^{(n)}$ in terms of $u_j^{(0)}$, we need to solve $n$ linear problems recursively. The procedure is straightforward to implement, but it will lead to increasingly complicated expressions as expected.\par

We have implemented our algorithm described here and applied it to the sum rule of OPE coefficient at one-loop order. For the example of $L=4,S=4,l=1,N=2$, we reproduce exactly \cite{Vieira:2013wya}
\begin{align}
F=\frac{16}{63}-\frac{196}{81}g^2\,.
\end{align}
Here, similar to the tree-level case, we can again use the
symmetrization trick in \eqref{symmetrization} to simplify the
Gr\"obner basis, companion matrix and the trace computation to get the
same answer.

Similar to the leading order, in practice it is more efficient to plug in the solution of BAE numerically than performing an analytic calculation on the quotient ring at higher loop orders. In particular, solving the linear problems numerically are much easier than solving it analytically. However, the analytic method gives exact results without possible loss of accuracy and avoids other subtleties of numerical methods. We should emphasis that our approach here is the most straightforward method, but not necessarily the most efficient one. There is still a huge room to improve the efficiency using the algebro-geometric methods.

\section{Conclusions, discussions and open questions}
\label{sec:conlcude}
In this paper we introduced the powerful language of computational algebraic geometry to study Bethe ansatz equations. We developed new analytical methods based on algebraic geometry to tackle two kinds of problems in the framework of Bethe ansatz.\par

To investigate the completeness problem of Bethe ansatz, we developed efficient methods to count the number of solutions of BAE for fixed quantum numbers with additional constraints. Our method is based on Gr\"obner basis and the quotient ring and are much faster than solving BAE numerically.\par

We developed an analytical method to perform the sum of any rational function over all physical solutions of BAE for fixed quantum numbers without actually solving the BAE. We applied our method to calculate the sum rules for OPE coefficients in $\mathcal{N}=4$ SYM both at tree level and one loop. We obtained exact rational numbers and proved that the results are always rational.\par

The most prominent advantage of our methods is the conceptual beauty. In the algebro-geometric language, solving BAE is equivalent to constructing the quotient ring of BAE. While the BAE can only be solved numerically, the quotient ring can be constructed analytically and systematically. The quotient ring is a finite dimensional linear space which can be studied much further. Any physical quantity in terms of rapidities is represented as finite dimensional matrix called companion matrix in the quotient ring. Eigenvalues of the companion matrix correspond to the values of this quantity at the solutions of BAE and trace of the companion matrix leads to the sum over all physical solutions at which the physical quantity are evaluated. In addition, constructing quotient rings of BAE is much more efficient than finding explicit solutions of BAE, which makes our method appealing also in practical applications.\par

There are many open problems that one can pursue in the near future. We list some of them below.

\begin{itemize}
\item As we mentioned in section\,\ref{sec:sum-rule-OPE}, Marboe and Volin recently proposed a new method to find physical solutions of BAE based on rational $Q$-systems \cite{Marboe:2016yyn,Marboe:2017dmb}. The analogy of BAE in this approach is the zero remainder conditions (ZRC) which are much easier to solve numerically. It is very interesting to combine our methods with the rational $Q$-systems and study the quotient ring of the ZRC. Our preliminary studies have shown that constructing the quotient ring of ZRC is much more efficient than constructing the quotient ring of BAE. It is also naturally much faster than solving the ZRC numerically. If the physical quantities we are interested in are symmetric with respect to the rapidities, we can study them equivalently in the quotient ring of ZRC. This will further boost the efficiency of computing the trace of the corresponding companion matrices.
\item Integrable models with symmetries of higher rank Lie algebras are solved by the so-called nested Bethe ansatz \cite{yang1967some,kulish:NBA,Belliard:ABA_all}, the resulting nested BAEs involve both physical rapidities and auxiliary ones and are hence much more complicated. It is an important future direction to investigate these more challenging cases using algebro-geometric methods. Furthermore, summing over physical quantities at all physical solutions of nested BAE has important applications in the recent work of asymptotic four-point functions \cite{Basso:2017khq}. The summand in the generalized sum rules in \cite{Basso:2017khq} is symmetric with respect to rapidities and we can apply the quotient ring of ZRC mentioned in the previous point.\par
\item A related question is how to construct companion matrices more efficiently. When the physical quantity under consideration is a complicated function of rapidities, the construction of the companion matrix can be quite tedious although straightforward. This is the main obstacle to the efficiency of our method. One possible way is to decompose the quantities into simpler parts. We can construct the corresponding companion matrices of the simpler parts and then combine them together. The latter step involves only operations on numerical matrices, which should be much easier to handle.
\item Concerning the completeness problem of the Heisenberg XXX spin chain, it is desirable to have analytic formula for the various numbers of solutions of BAE with different constraints in terms of $L$ and $N$. There are several relevant theorems for this type of counting such as the B\'ezout theorem and the more refined BKK theorem. Using these theorems in an ingenious way and combining some possible local analysis for the special cases, this ambitious goal does not seem to be impossible.
\item It is also interesting to investigate the completeness problem of Heisenberg spin chains in higher spin representations where a similar conjecture like the one in (\ref{eq:conjecture_numbers}) has been proposed \cite{Hao:2013rza}. The generalization to the cases with different boundary conditions \cite{sklyanin1988boundary,Nepomechie:2003ez,wang2015off:book,cao2013off:XXX} is also an interesting problem.
\item One particularly interesting direction is to generalize our current method to the quantum deformed XXZ spin chain. In this case, we have an additional parameter, namely the anisotropy to play with. The completeness problem was investigated in \cite{langlands1995algebro} (see also \cite{Brattain} for the generalization to the case of BAE in the asymmetric simple exclusion processes.) It is known from numerics\footnote{For the two magnon case, a much more thorough analysis can be performed using the properties of self-inversive polynomials \cite{Vieira,Vieira:numberofroots}.} that the structures of solutions of BAE are different in different regimes of anisotropy (see for example \cite{Fabricius2001,Nepomechie:2003ez}). It will be fascinating to see this kind of change in the structure of the quotient rings.
\item Finally, we only applied the technique of Gr\"obner basis and resultants to two kinds of problems in the current paper. There are many other powerful tools in algebraic geometry as well as many interesting problmes in integrable models. A wider range of applications and a deeper mutual fertilization could be expected. One particularly interesting example is the computation of \emph{exact} partition function of integrable lattice models such as six vertex model. A related question is computing grand partition functions of $\mathcal{N}=4$ SYM theory at one-loop \cite{Suzuki:2017ipd}.
\end{itemize}

\section*{Acknowledgements}
We acknowledge N. Beisert, J. Boehm, C. Eder, H.
Ita, K. Larsen and H. Sch\"onemann for enlightened
discussions. Especially, we thank C. Eder for testing the
Gr\"obner basis computation for BAE with the package 'gb'.
Y. Jiang and Y. Zhang are partially supported by the Swiss National
Science Foundation through the NCCR SwissMap. Y. Zhang's research leading to these results received funding from
Swiss National Science Foundation (Ambizione grant
PZ00P2\_161341).

\appendix

\section{More on completeness of BAE}
\label{app:BAEcompleteness}
In this appendix, we discuss the four kinds of special solutions in more detail. The discussions below require some basic knowledge about algebraic Bethe ansatz, for which we refer to \cite{Faddeev:ABA}.

\paragraph{Coinciding rapidities}
If we solve BAE without any constraints, we indeed find solutions of the form $\{u,u,u_1,\cdots,u_N\}$. They are legitimate solutions of BAE. In the case of coinciding roots, the BAE take a slightly different from which we derive below. Let us recall the $RTT$ relation for XXX$_{1/2}$ spin chain
\begin{align}
A(\lambda)B(\mu)=&\,f(\lambda,\mu)B(\mu)A(\lambda)+g(\lambda,\mu)B(\lambda)A(\mu),\\\nonumber
D(\lambda)B(\mu)=&\,f(\mu,\lambda)B(\mu)D(\lambda)+g(\mu,\lambda)B(\lambda)D(\mu),\\\nonumber
B(\lambda)B(\mu)=&\,B(\mu)B(\lambda).
\end{align}
Using these relations, one can derive the following result \cite{Izergin1982,Avdeev1986}
\begin{align}
\label{eq:ABB}
A(\lambda)B(\mu)^2=&\,a_1(\lambda,\mu)B(\mu)^2 A(\lambda)+a_2(\lambda,\mu)B(\lambda)\textcolor{blue}{B'(\mu)}A(\mu)\\\nonumber
&\,+a_3(\lambda,\mu)B(\lambda)B(\mu)A(\mu)+a_4(\lambda,\mu)B(\lambda)B(\mu)\textcolor{blue}{A'(\mu)}\\\nonumber
D(\lambda)B(\mu)^2=&\,b_1(\lambda,\mu)B(\mu)^2 D(\lambda)+b_2(\lambda,\mu)B(\lambda)\textcolor{blue}{B'(\mu)}D(\mu)\\\nonumber
&\,+b_3(\lambda,\mu)B(\lambda)B(\mu)A(\mu)+b_4(\lambda,\mu)B(\lambda)B(\mu)\textcolor{blue}{D'(\mu)}
\end{align}
where $a_i(\lambda,\mu)$ and $b_i(\lambda,\mu)$ $(i=1,2,3,4)$ are some functions of $f(\mu,\lambda)$ and $g(\mu,\lambda)$. As we can see, due to the presence of coinciding rapidities, we have operators $B'(u)=\partial_u B(u)$ as well as $A'(u)=\partial_u A(u)$ and $D'(u)=\partial_u D(u)$. In order the off-shell Bethe state
\begin{align}
|\Psi\rangle=B(u)^2\prod_{i=1}^N B(u_i)|\uparrow^L\rangle
\end{align}
be an eigenstate of the transfer matrix, one computes
\begin{align}
T(v)|\Psi\rangle=(A(v)+D(v))|\Psi\rangle
\end{align}
by moving the diagonal elements $A(u)$ and $D(u)$\footnote{In the case of coinciding rapidities, we also move the corresponding operators with derivatives $A'(u)$ and $D'(u)$ to the rightmost.} to the rightmost and acting on the pseudovacuum using the commutation relations (\ref{eq:ABB}). This will generate the so-called `wanted terms' and `unwanted terms'. By demanding the unwanted terms to vanish, one obtains the usual BAE. There are two modifications in the current case. First of all, the appearance of $A'(v)$ and $D'(v)$ lead to $a'(u)$ and $d'(u)$ which might modify the form of the cancelation conditions. In addition, the appearance of $B'(u)$ means we need to impose cancelation conditions for the states involving $B'(u)$. If we have more coinciding rapidities, from similar analysis we have more additional cancelation conditions.\par

In fact the cancelation conditions can be obtained most easily by demanding that the eigenvalue of transfer matrix is regular at the Bethe roots. Consider the solution BAE of a spin chain of length $L$ in the spin-$s$ representation with $K+N$ magnons $\{u,u,\cdots,u,u_1,\cdots,u_N\}$. The eigenvalue of the transfer matrix is given by
\begin{align}
T(\lambda)=a(\lambda)\left(\frac{\lambda-u-i}{\lambda-u}\right)^K\prod_{j=1}^N\frac{\lambda-u_j-i}{\lambda-u_j}+
d(\lambda)\left(\frac{\lambda-u+i}{\lambda-u}\right)^K\prod_{j=1}^N\frac{\lambda-u_j+i}{\lambda-u_j}
\end{align}
where
\begin{align}
a(\lambda)=(\lambda+is)^L,\qquad d(\lambda)=(\lambda-is)^L.
\end{align}
By construction, $T(\lambda)$ is a polynomial in $\lambda$ although it seems to have poles at $\lambda=u,u_1,\cdots,u_N$. By requiring the residues of these `poles' to vanish, we obtain the BAE. For $\mu=u_j$, $(j=1,\cdots,N)$ we have
\begin{align}
\rB_j=a(u_j)(u_j-u-i)^KQ_{\mathbf{u}}(u_j-i)+d(u_j)(u_j-u+i)^KQ_{\mathbf{u}}(u_j+i)=0
\end{align}
Requiring $\lambda=u$ is regular leads to the following conditions
\begin{align}
\rR_l=\frac{\partial^l}{\partial\lambda^l}\left.\left(T(\lambda)(\lambda-u)^K  \right)\right|_{\lambda=u}=0,\qquad l=0,\cdots,K-1.
\end{align}
It was proved in \cite{Izergin1982} that for the 1D Bose gas where $a(u)=e^{-iuL}, d(u)=e^{+iuL}$, the BAE $\rB_j=\rR_l=0$ do not have solutions for $K\ge 2$. For the Heisenberg spin chain, it was found in \cite{Avdeev1986} that there are no solutions with $K\ge 3$ and the ones with more than one group of repeated roots such as $\{u,u,v,v,u_1,\cdots,u_N\}$. However, one can find many solutions of the form $\{u,u,u_1,\cdots,u_N\}$. Therefore, apart from the general believe that these solutions are not physical, there is no rigorous mathematical proof to this assertion as in the case of 1D Bose gas.

\paragraph{Solutions beyond the equator}
When looking for physical solutions, we usually restrict ourselves to the regime $N\le L/2$. The BAE itself is well defined also for $N>L/2$ and explicit solutions can be found. Why do we neglect these solutions ? The answer is that they are already included in the first case. To understand this, let us consider the $N<L/2$ magnon Bethe state of a spin chain of length $L$. The Bethe vector can be generated by acting $N$ operators $B(u)$ on the pseudovacuum
\begin{align}
\label{eq:state1}
|\Psi\rangle=B(u_1)\cdots B(u_N)|\uparrow^L\rangle
\end{align}
where the rapidities should satisfy the BAE of  $N$ particles. This state has $N$ down spins and $L-N$ up spins. We can generate the eigenstate with the same amount of up spins and down spins by acting $L-N$ operators $C(v)$ on the flipped pseudovacuum
\begin{align}
\label{eq:state2}
|\tilde{\Psi}\rangle=C(v_1)\cdots C(v_{L-N})|\downarrow^L\rangle
\end{align}
Now the rapidities $v_1,\cdots,v_{L-N}$ should satisfy the BAE of  $L-N$ particles. As it turns out $|\Psi\rangle=|\tilde{\Psi}\rangle$, so (\ref{eq:state1}) and (\ref{eq:state1}) are merely two ways of constructing the same eigenstate. It is then clear that $\mathbf{u}=\{u_1,\cdots,u_N\}$ and $\mathbf{v}=\{v_1,\cdots,v_{L-N}\}$ should be related. This is indeed the case. To see this, one can define the Baxter polynomials
\begin{align}
Q_{\mathbf{u}}(u)=\prod_{k=1}^N (u-u_k),\qquad Q_{\mathbf{v}}(u)=\prod_{k=1}^{L-N}(u-v_k).
\end{align}
It can be shown that the two polynomials satisfy the Wronskian relation, which implies that knowing one of the polynomials gives us the other one. The two polynomials are in fact two solutions of Baxter's $TQ$-relation which is a second order difference equation. The above analysis shows that we can safely restrict ourselves to one side of the equator $N\le L/2$. The other solutions lead to the same physical states.

\paragraph{Solutions at infinity}
The Bethe states which correspond to rapidities $\{u_1,u_2,\cdots,u_N\}$ with none of the elements at infinity is the so-called highest weight state. This means
\begin{align}
S^+ B(u_1)B(u_2)\cdots B(u_N)|\Omega\rangle=0,\qquad S^+=\sum_{i=1}^L S_i^+.
\end{align}
The above relation is non-trivial but can be proved rather straightforwardly. The corresponding spin of this highest weight state is $J=\frac{L}{2}-N$. As in quantum mechanics, we can use $S^-$ to lower the spins. For a spin-$J$ representation, the dimension is $2J+1$. Therefore, for a highest weight state $|u_1,\cdots,u_N\rangle$, the following states
\begin{align}
(S^-)^n|u_1,\cdots,u_N\rangle,\qquad n=0,\cdots, L-2N
\end{align}
form a representation space of $\mathfrak{su}(2)$ algebra. For the completeness of Bethe ansatz, it is thus expected that the number of physical solutions of $N$-particle BAE should be
\begin{align}
Z_{L,N}={L\choose N}-{L\choose N-1}
\end{align}
Then the total number of Bethe states is
\begin{align}
\sum_{N=0}^{L/2}Z_{L,N}(L-2N+1)=2^L
\end{align}
which is the dimension of the Hilbert space. The solution of BAE allows putting one or more excitations to infinity. Each rapidity at infinity correspond to acting an $S^-$ due to the fact
\begin{align}
\lim_{u\to \infty}B(u)\propto S^-.
\end{align}
Therefore solutions at infinity are allowed and are physical. To show the completeness of Bethe ansatz, we only need to count the solutions that correspond to primary states, the descendants of a primary state is easy to work out. Therefore when we count the solutions, we only count the ones corresponding to primary states.

\paragraph{Singular solutions}
The solutions of BAE with two of the rapidities being $\pm i/2$, namely
\begin{align}
\{i/2,-i/2,u_3,\cdots,u_N\}
\end{align}
are called singular solutions. To see that there is a problem at $u=\pm i/2$, it is simplest to look at the eigenvalue in terms of the rapidities
\begin{align}
E_N=-\frac{1}{2}\sum_{k=1}^N\frac{1}{u_k^2+1/4}.
\end{align}
It is obvious that the function $(u^2+1/4)^{-1}$ have two poles located at $u=\pm i/2$. Therefore solutions containing $u=\pm i/2$ are special. These solutions are more subtle than the ones we discussed before. The reason is that sometimes these solutions are physical and sometimes not. To see whether a solution is physical or not, one needs to perform a judicious regularization. Such analysis has been worked out in detail in the work of Nepomechie and Wang \cite{Nepomechie:2013mua}. The conclusion of their analysis is that the solutions are physical if the remaining rapidities $u_3,\cdots,u_N$ satisfy the following equations
\begin{align}
\label{eq:phys_sings}
\left(\frac{u_k+i/2}{u_k-i/2}\right)^{L-1}\left( \frac{u_k-3i/2}{u_k+3i/2}\right)=&\,\prod_{j\ne k\atop j=3}^M\frac{u_k-u_j+i}{u_k-u_j-i},\qquad k=3,\cdots,N.\\\nonumber
\prod_{k=3}^N\left(\frac{u_k+i/2}{u_k-i/2}\right)^L=&\,(-1)^L.
\end{align}
The first equation is the usual BAE while the second one is an additional selection rule.

\section{OPE coefficients and sum rules in $\mathcal{N}=4$ SYM}
\label{sec:OPE}
In this appendix, we give more details about the OPE coefficients and sum rules in the main text. We mainly follow the discussion in \cite{Vieira:2013wya}. The OPE coefficients can be obtained by computing three-point functions. In our case, we need to compute the three-point function with two BPS operators and one non-BPS operator in the $SL(2)$ sector.\par

The three operators under consideration are the following. First we have two BPS operators which takes the following form
\begin{align}
\mathcal{O}_1^{\text{BPS}}(x_1)=&\,\tr(\bar{Z}\bar{X}\bar{X}\bar{Z}\cdots)(x_1)+\cdots\\\nonumber
\mathcal{O}_2^{\text{BPS}}(x_2)=&\,\tr(ZZXX\cdots)(x_2)+\cdots
\end{align}
where $Z$ and $X$ are two complex scalar fields and $\bar{Z}$, $\bar{X}$ are the corresponding complex conjugates. The third operator is a non-BPS and takes the following form
\begin{align}
\mathcal{O}_3^S(x_3)=\sum_{1\le n_1\le n_2\le\cdots\le n_S\le L}\psi(n_1,n_2,\cdots,n_S)\mathcal{O}_{n_1,n_2,\cdots,n_S}(x_3).
\end{align}
The wave functions $\psi(n_1,n_2,\cdots,n_S)$ depend on the Bethe roots, namely the solution of Bethe ansatz equations. The operators $\mathcal{O}_{n_1,n_2,\cdots,n_S}$ are given by
\begin{align}
\mathcal{O}_{n_1,n_2,\cdots,n_S}=\left[\prod_{j=1}^{L}\frac{1}{m_j!}\right]\tr\left(Z\cdots Z\underset{n_1}{D}Z\cdots\underset{n_2}{D}Z\cdots  \right)
\end{align}
where $D$ is the covariant derivative projected to some light-cone direction $D=D_\mu n^\mu$ with $n^2=0$.\par

For the two BPS operators, the lengths of the operators are defined as the total number of the scalar fields. We denote the lengths of BPS operators to be $L_1$ and $L_2$ and the number of scalar fields $X$ (which is equal to the number of scalar fields of $\bar{X}$) to be $N$. We also define $l=L_1-N$, which is the number of scalar field $Z$ for operator $\mathcal{O}_1$. Let us denote the length (sometimes called twist, which is the number of scalar fields) of the non-BPS operator to be $L_3=L$ and the total number of covariant derivatives as $S$. Then we have the following relation
\begin{align}
L_1=l+N,\qquad L_2=N+L-l,\qquad L_3=L.
\end{align}
and the number of covariant derivatives of $\mathcal{O}_3$ is $S$, which is also the number of Bethe roots.\par

The three-point functions of the three operators which we describe above is completely fixed up to a constant called the structure constant, which is the OPE coefficient that appears in the sum rule.
\begin{align}
\langle\mathcal{O}_1^{\text{BPS}}(x_1)\mathcal{O}_2^{\text{BPS}}(x_2)\mathcal{O}_3^S(x_3)\rangle
=\frac{1}{N_c}\frac{{C_{\mathbf{u}}^{\bullet\circ\circ}}}{x_{12}^{\Delta-S+2l-L}x_{13}^{\Delta-S+L-2l}x_{23}^{L+N-(\Delta-S)}}\left(\frac{x_{12}^\mu n_\mu}{x_{12}^2}-\frac{x_{13}^\mu n_\mu}{x_{13}^2}\right)^S
\end{align}
The explicit expression of $C_{\mathbf{u}}^{\bullet\circ\circ}$ is given in (\ref{eq:OPEcoe_explicit}), (\ref{eq:OPEcoe_A}) and (\ref{eq:OPEcoe_B}). The non-perturbative expression of the momentum and $S$-matrix are given by
\begin{align}
e^{ip(u_j)}=\frac{x_j^+}{x_j^-},\qquad S(u_j,u_k)=\frac{u_j-u_k+i}{u_j-u_k-i}\left(\frac{1-1/x_j^-x_k^+}{1-1/x_j^+x_k^-}\right)^2\sigma(u_j,u_k)^2
\end{align}
where
\begin{align}
x_j^\pm\equiv x(u_j\pm i/2),\qquad x(u)=\frac{u+\sqrt{u^2-4g^2}}{2g}
\end{align}
and $\sigma(u_j,u_k)$ is the so-called BES dressing phase \cite{Beisert:2004hm}. The dressing phase is a rather complicated quantity but it will only start to contribute at three-loops.\par

We define and expand the sum rule as the follows
\begin{align}
\sum_{\text{sol. fixed $L$ and $S$}}\left(C_{\mathbf{u}}^{\bullet\circ\circ}\right)^2 e^{\gamma_{\mathbf{u}}y}=\sum_{n=0}^\infty g^{2n}\sum_{m=0}^n y^m\,\mathcal{P}_S^{(n,m)}
\end{align}
where $y$ is an auxiliary variable. By computing the sum rule, one has predictions for the numbers $\mathcal{P}_S^{(n,m)}$, which can also be obtained from four-point functions in the OPE limit. For more details, we refer to \cite{Vieira:2013wya}. From the four-point function side, it is clear that $\mathcal{P}_S^{(m,n)}$ are rational numbers. By comparing the numbers $\mathcal{P}_S^{(n,m)}$ from different approaches, one can check the validity of the integrability-based calculations.

\section{Method of resultant}
\label{sec:resultant}
In this appendix, we introduce another method to count the number of solutions of BAE with additional constraints. This method avoids the computation of Gr\"obner basis and uses another important object of computational algebraic geometry, which is the \emph{resultant}.\par

Recall that the multi-variable resultant of the \emph{homogeneous}
polynomials $F_0,\cdots,F_n\in\mathbb{C}[x_0,\cdots,x_n]$  is a
uniquely defined polynomial in terms of coefficients of the
coefficients of $F_i$ with the crucial property that whenever the
equations $F_0=\cdots=F_n=0$ has a non-trivial solution, the so-called {\it
  Macaulay resultant} $\text{Res}(F_0,\cdots,F_n)=0$
\cite{opac-b1094391}. Our method is based on this fundamental property.\par

Suppose we have to solve $n$ polynomial equations given by $f_1=\cdots
f_n=0$ where $f_i\in\mathbb{C}[u_1,\cdots,u_N]$. The polynomials
$f_i(u_1,\cdots,u_n)$ are not necessarily homogeneous. We then pick
one of the variables, say $u_1$ (We can pick any $u_k$) and view it as
a \emph{parameter}. Then $f_i$ are polynomials depending on variables
$u_2,\cdots,u_n$. In order to define the resultant, we introduce
another variable $u_0$ to homogenize the polynomials. Let us denote
the homogenized polynomials by $F_i(U_0,U_2,\cdots,U_n;
u_1)$\footnote{We use capital letters to denote the variables and
  lower case ones to denote parameters, where $U_i/U_0=u_i$},
$(i=1,\cdots,n)$ and we have
$F_i(1,u_2,\cdots,u_n;u_1)=f_i(u_1,u_2,\cdots,u_n;u_1)$. We can then compute
the resultant of the polynomials $F_i(U_0,U_2,\cdots,U_n;u_1)$ which is
now a polynomial depending on $u_1$. We then have
\begin{align}
q(u_1)=\text{Res}(F_1,\cdots,F_n).
\end{align}
The claim is that the number of solutions for the single variable polynomial $q(u_1)=0$, or equivalently, the highest power of the polynomial $q(u_1)$ gives the number of solutions for the original equations $f_1=\cdots=f_n=0$. \footnote{Note that the original Macaulay
  resultant computation requires the number of equations equals the
  number of variables. In practice, we may have the situations for
  which the number of equations is larger then the number of
  variables. In these cases, the idea of Macaulay can also apply
  through the evaluation of several Macaulay resultants. For example,
  suppose that we have $n+1$ equations $f_1=\ldots =f_{n+1}=0$ in $n$
  variables. With the same notations, we can homogenize the variables
  except $u_1$ and get $n+1$ homogeneous polynomials $F_i(U_0, U_2,
  \ldots U_n; u_1)$, $i=1,\ldots, n+1$. Then we calculate two
  resultants,
  \begin{equation}
    \label{eq:13}
    q(u_1)=\text{Res}(F_1,\cdots,F_n),\quad p(u_1)=\text{Res}(F_2,\cdots,F_{n+1}).
  \end{equation}
Eventually, we calculate the greatest common factor, $\gcd(q,p)$ of $q(u_1)$ and
$p(u_1)$. The high power of $\gcd(q,p)$ provides the number of solutions the number of solutions for the original equations $f_1=\cdots=f_{n+1}=0$
  }\par

Let us illustrate our general procedure by a simple example. We consider the following equations $f_1=f_2=f_3=0$ where $f_i(u_1,u_2,u_3)$ is given by
\begin{align}
f_1=&\,u_1^2+u_2^2+u_3^2-3,\\\nonumber
f_2=&\,u_1^2+u_3^2-2,\\\nonumber
f_3=&\,u_1^2+u_2^2-2u_3.
\end{align}
We view $u_3$ as a parameter and introduce another variable $u_0$ to homogenize the three polynomials, which leads to three homogenized polynomials $F_i(U_0,U_1,U_2;u_3)$, $(i=1,2,3)$
\begin{align}
F_1=&\,U_1^2+U_2^2+(u_3^2-3)U_0^2,\\\nonumber
F_2=&\,U_1^2+(u_3^2-2)U_0^2,\\\nonumber
F_3=&\,U_1^2+U_2^2-2u_3\,U_0^2.
\end{align}
The resultant of $F_1,F_2,F_3$ is now a polynomial in $u_3$
\begin{align}
q(u_3)=\text{Res}(F_1,F_2,F_3).
\end{align}
The resultant $\text{Res}(F_1,F_2,F_3)=0$ if and only if there is a non-trivial solution $(U_0,U_2,U_3)\ne(0,0,0)$ of the equation $F_1=F_2=F_3=0$. The resultant can be evaluated explicitly
\begin{align}
\label{eq:resultant_u3}
q(u_3)=(u_3^2+2u_3-3)^4.
\end{align}
Suppose we find a root of $q(u_3)=0$, denoted by $\bar{u}_3$. Then for $u_3=\bar{u}_3$, the equations $F_1=F_2=F_3=0$ have non-trivial solutions, which we denote by $(\overline{U}_0,\overline{U}_1,\overline{U}_2)$. The solution is projective. That is to say for fixed $\bar{u}_3$, if $(\overline{U}_0,\overline{U}_1,\overline{U}_2)$ is a non-trivial solution, then for any $\lambda\ne 0$, $(\lambda\overline{U}_0,\lambda\overline{U}_1,\lambda\overline{U}_2)$ is also a non-trivial solution. We can use this freedom to rescale $\overline{U}_0$ to be 1 and denote the corresponding solution as $(1,\bar{u}_1,\bar{u}_2)$. It is then clear that $(\bar{u}_1,\bar{u}_2,\bar{u}_3)$ is the solution of the original equations $f_1=f_2=f_3=0$. Therefore each solution of $q(u_3)=0$ corresponds to a solution of the original equations. Since $q_3(u_3)$ is a polynomial of a single variable, the number of solution is simply the highest power of $q(u_3)$. For our current example, we find immediately from (\ref{eq:resultant_u3}) that the number of solutions is 8. This is in agreement with a direct solution $(-1,\pm1,1),(1,\pm1,1),(\sqrt{7}i,\pm -3),(-\sqrt{7}i,\pm1,-3)$.\par

The main computation in this approach is the multi-variable Macaulay
resultant. {We find that so far the resultant
  computation for BAE is complicated and not as efficient as the
  Gr\"obner basis method. Since the resultant is given in terms of
  \emph{determinants} of \emph{large} sparse matrices, we expect that
  in the future,
  the special Gaussian elimination method optimized for Macaulay
  matrix can speed up the resultant computation drematically, and make
this method applicable for complicated BAE. (For example, the
GBLA algorithm described in \cite{DBLP:journals/corr/BoyerEFLM16} has a simple method
of reducing large Macaulay
matrices. However, the specific function for computing Macaulay
resultant via GBLA algorithm is not available to the public yet.)}


\section{Computation of Gr\"obner basis}
\label{sec:Gr}
A Gr\"obner
basis can be computed by various algorithms like Buchberger \cite{Buchberger:1976:TBR:1088216.1088219}, F4
\cite{FAUGERE199961} or F5 \cite{Faugere:2002:NEA:780506.780516}
algorithms. The classical Buchberger algorithm is the simplest (but
may not be the most efficient) algorithm. To provide some intuitions
of Gr\"obner basis computations, in this appendix we first briefly
review Buchberger algorithm.

Given two polynomials $f$ and $g$ in a polynomial ring $\mathbb K[x_1 , \ldots x_n]$
with a monomial order $\succ$, we can define the S-polynomials of $f$ and $g$ as,
\begin{equation}
  \label{eq:10}
  S(f,g) \equiv \frac{\text{LCM}(\LT(f), \LT(g))}{\LT(f)} f -\frac{\text{LCM}(\LT(f),
    \LT(g))}{\LT(g)} g\,.
\end{equation}
Here $\text{LCM}$ means the least common multiplier, and $\LT$ means
the leading term of a polynomial in the given monomial order. It is
clearly that $S(f,g)$ is a polynomial generated by $f$ and $g$.

Given a polynomial set $\{f_1, \ldots f_k\}$ in $\mathbb K[x_1 ,
\ldots x_n]$, the G\"obner basis can be computed by Buchberger
algorithm as follows:
\begin{enumerate}
\item Create a list $B=\{f_1, \ldots f_k\}$ and a queue $l$ of all polynomial
  pairs in $B$, $(f_i,f_j)$, $i\leq j$.
\item Pick up the head of the queue, say, $(f,g)$. Calculate the
  S-polynomial $S(f,g)$. Divide $S(f,g)$ towards $B$ and get the
  reminder $r$. Delete the head of the queue $l$.
\item If $r$ is non-zero, add $r$ to the list $B$ and also add
  polynomials pairs consisting of $r$ and elements in $B$ to the queue
  $l$.
\item If the queue $l$ is empty, the list $B$ is required Gr\"obner
  basis and the algorithm stops. Otherwise, go to step 2.
\end{enumerate}
To illustrate this algorithm, we can compute a simple Gr\"obner basis
\cite{MR3330490}.  Consider $f_1=x^3 - 2 x y$, $f_2=x^2 y - 2 y^2 +
x$. Compute the G\"obner basis of
  $I=\langle f_1,f_2\rangle$ with the DegreeReverseLexicographic order and
  $x\succ y$:
\begin{enumerate}
\item In the beginning, the list is $B=\{h_1,h_2\}$ and the queue is
  $l=\{(h_1,h_2)\}$, where
$h_1=f_1$, $h_2=f_2$,
\begin{equation}
  S(h_1,h_2)=-x^2,\quad h_3=\overline{S(h_1,h_2)}^B=-x^2\,,
\end{equation}
Here $\overline{S(h_1,h_2)}^B$ means the remainder of the S-polynomial
$S(h_1,h_2)$ from its division towards $B$.
\item Now $B=\{h_1,h_2,h_3\}$ and
$l=\{(h_1,h_3),(h_2,h_3)\}$. Consider the pair $(h_1,h_3)$,
\begin{equation}
  S(h_1,h_3)=2xy,\quad h_4=\overline{S(h_1,h_3)}^B=2 x y\,,
\end{equation}
\item $B=\{h_1,h_2,h_3,h_4\}$ and $l=\{(h_2,h_3),(h_1,h_4),(h_2,h_4),(h_3,h_4)\}$.
For the pair $(h_2,h_3)$,
\begin{equation}
  S(h_2,h_3)=-x+2y^2,\quad h_5:=\overline{S(h_2,h_3)}^B=-x+2y^2\,,
\end{equation}
\item $B=\{h_1,h_2,h_3,h_4,h_5\}$ and
\begin{equation}
  l=\{(h_1,h_4),(h_2,h_4),(h_3,h_4),(h_1,h_5),(h_2,h_5),(h_3,h_5),(h_4,h_5)\}.
\end{equation}
For the pair $(h_1,h_4)$,
\begin{equation}
  S(h_1,h_4)=-4 x y^2,\quad \overline{S(h_1,h_4)}^B=0
\end{equation}
Hence this pair does not add a new polynomial to the basis. Similarly, all the
rests pairs contain no new information.
\end{enumerate}
Therefore the Gr\"obner basis is
\begin{equation}
  \label{eq:42}
  B=\{h_1,\ldots h_5\}=\{x^3-2 x y,x^2 y+x-2 y^2,-x^2,2 x y,2 y^2-x\}.
\end{equation}

So far this algorithm provides the Gr\"obner basis in the given
monomial ordering. In many cases, this kind of Gr\"obner bases are
enough for the practice. However, it is not in the ``simplest form'',
the {\it minimal reduced} Gr\"obner basis. A minimal reduced Gr\"obner basis is a
Gr\"obner basis such that the leading term from any polynomial in the
basis cannot
divide any monomial in other polynomials in this basis. The minimal reduced Gr\"obner basis
of an ideal is {\it unique} for a given monomial ordering.

To determine the minimal reduced  Gr\"obner basis for this example, we can do
the following: note that $\LT(h_3)|\LT(h_1)$,
$\LT(h_4)|\LT(h_2)$, so $h_1$ and $h_2$ are removed from the basis. Furthermore,
\begin{equation}
  \label{eq:54}
  \overline{h_3}^{\{h_4,h_5\}} =h_3,\quad \overline{h_4}^{\{h_3,h_5\}}
  =h_4, \quad \overline{h_5}^{\{h_3,h_4\}} =h_5\quad
\end{equation}
so $\{h_3,h_4,h_5\}$ cannot be reduced further. The minimal reduced Gr\"obner basis is
\begin{equation}
  \label{eq:56}
  G=\{-x^2,2 x y,2 y^2-x\}.
\end{equation}

Buchburger algorithm is simple and intuitive. However, it
requires the reduction of many polynomials pairs and can be slow in
the practice. In this paper, we mainly used 'slimgb'  in the software
\rm{Singular} \cite{DGPS} and the C library 'gb'
\cite{GB,DBLP:journals/corr/BoyerEFLM16} and package 'FGb' \cite{FGb} for computing Gr\"obner
bases:
\begin{itemize}
\item 'slimgb' is an improved Buchberger algorithm
  \cite{Brickenstein2010} which smartly picks
  up the polynomial pairs to reduce the size of intermediate results
  to speed up the computation.
\item 'Fgb' is a modern Gr\"obner
basis package written by Jean-Charles Faug\`ere which applies F4 and
F5 algorithm. It can reduce a lot of S-pairs at once and automatically
drop the useless S-pairs in the computation.
\item 'gb' is a new Gr\"obner
basis C Library written by Christian Eder based on the GBLA
algorithm \cite{DBLP:journals/corr/BoyerEFLM16} and fast linear
algebra techniques.
\end{itemize}

Sometimes, the Gr\"obner basis computation over $\mathbb Q$ is
slow. In this case, we can first calculate the Gr\"obner basis over
finite fields $Z/p_1, \ldots , Z/p_k$, where $p_1, \ldots , p_k$ are
prime numbers. Then we can use Chinese remainder theorem and Farey
fractions to lift the finite-field results to rational results. The
step can be automated by the package 'modstd\_lib' in \rm{Singular}.



\section{\texttt{Maple} and \texttt{Mathematica} codes}
\label{sec:codes}

We attach ``L12M5.mw'', the Maple file for computing the Gr\"obner
basis for nonsingular Bethe roots in $SU(2)$ model with $L=12$ and
$N=5$ with 'FGb' package, and also ``SL\_L4S4.nb'', the Mathematica file
for computing structure constant in $SL(2)$ model with $L=4$ and
$S=4$, as computation examples.

To run ``L12M5.mw'', it is necessary to install 'Fgb' package for
Maple first. To run ``SL\_L4S4.nb'', it is necessary to install {\rm
  Singular} and furthermore to download the {\rm Mathematica} packages
for {\rm Singular} interface and quotient ring basis
computations. These {\rm Mathematica} packages are included in this submission.

\bibliographystyle{JHEP}
\bibliography{yunfeng}

\end{document}